\documentclass[twocolumn]{aastex63}

\usepackage{amsmath}
\usepackage{ulem}
\usepackage{xcolor}   
\usepackage{url}

\newcommand{\degrees}{\ensuremath{^\circ}}

\newcommand{\new}[1]{{\bf #1}}

\newcommand\tabblends{
\begin{table*}
\caption{Prominent lines 
and blends in the  solar chromospheric spectrum}
\label{tab:blends}
\scriptsize
\begin{center}
\begin{tabular}{llllllll}
\hline
\hline
$\lambda$ nm & Land\'e g & log $\omega_L$ & log $\sum_iA_{ji}$ & $B_{HANLE}$ & log $\tau_0$ & Ion & blend \\
lab. (air) & factor & rad~sec$^{-1}$ & sec$^{-1}$& G & & & severity$^\ast$ Ion $\lambda$\\
\hline
256.254 &  1.21 &  7.03 &  8.49 &   28.8 &  -2.54&Fe II&  \\
 256.348  & 1.10  & 6.99  & 8.49  &  31.9  & -2.87 &Fe II &1 Fe I .341\\
 256.691  & 0.83  & 6.86  & 8.49  &  42.2  & -3.30 &Fe II &  \\
 257.437  & 1.30  & 7.06  & 8.51  &  28.4  & -4.33 &Fe II &1 Co I .351\\
 257.792  & 1.33  & 7.07  & 8.49  &  26.4  & -3.32 &Fe II &1 Mg I .888\\
 258.258  & 1.47  & 7.11  & 8.49  &  23.9  & -3.19 &Fe II &  \\
 258.588  & 1.50  & 7.12  & 8.46  &  21.7  & -2.03 &Fe II &2 Fe I .588\\
 259.055  & 1.50  & 7.12  & 8.48  &  22.8  & -4.47 &Fe II &1 Co I .059\\
 259.154  & 1.49  & 7.12  & 8.49  &  23.5  & -3.24 &Fe II &  \\
 259.373  & 2.17  & 7.28  & 8.49  &  16.2  & -4.02 &Fe II &1 Mn I .372\\
 259.837  & 1.50  & 7.12  & 8.46  &  21.7  & -1.93 &Fe II &  \\
 259.940  & 1.56  & 7.14  & 8.46  &  21.1  & -1.39 &Fe II &1 Fe I .957\\
260.709  & 1.50  & 7.12  & 8.66  &  34.8  & -1.99 &Fe II &  \\
 261.107  & 1.90  & 7.22  & 8.49  &  18.4  & -4.23 &Fe II &1 Cr II? .104\\
 261.187  & 1.59  & 7.15  & 8.46  &  20.5  & -1.79 &Fe II &  \\
 261.382  & 1.50  & 7.12  & 8.46  &  21.7  & -2.20 &Fe II &  \\
 261.762  & 1.66  & 7.16  & 8.46  &  19.6  & -2.36 &Fe II &  \\
 262.041  & 1.87  & 7.22  & 8.66  &  27.9  & -3.75 &Fe II &  \\
 262.167  & 3.34  & 7.47  & 8.46  &   9.8  & -2.76 &Fe II &  \\
 262.567  & 1.50  & 7.12  & 8.46  &  21.9  & -2.08 &Fe II &  \\
 262.829  & 1.50  & 7.12  & 8.66  &  34.8  & -2.21 &Fe II & \\
 263.105  & 1.50  & 7.12  & 8.46  &  21.7  & -2.02 &Fe II &1 Fe II .045$^\dag$\\
 263.132  & 1.50  & 7.12  & 8.46  &  21.7  & -1.97 &Fe II &  \\
 264.112  & 1.87  & 7.22  & 8.51  &  19.8  & -4.62 &Fe II &1 Ti I .089\\
 268.351  & 1.84  & 7.21  & 8.51  &  20.1  & -4.99 &Fe II &2 Cr II .345\\
 269.283  & 1.93  & 7.23  & 8.47  &  17.2  & -4.40 &Fe II &1 Fe I .265, Fe II .260\\
 270.938  & 2.10  & 7.27  & 8.47  &  15.8  & -4.30 &Fe II &1 Cr II .931\\
 271.441  & 1.50  & 7.12  & 8.60  &  30.3  & -3.09 &Fe II &  \\
 271.670  & 1.33  & 7.07  & 8.46  &  24.8  & -3.11 &Fe II &1 Mn II .680\\
 271.903  & 1.25  & 7.04  & 8.25  &  16.3  & -4.45 &Fe  I &2 Fe I .903\\
 272.090  & 1.17  & 7.01  & 8.28  &  18.4  & -4.76 &Fe  I &  \\
 272.358  & 1.00  & 6.94  & 8.27  &  21.1  & -5.22 &Fe  I &  \\
 272.488  & 1.20  & 7.02  & 8.47  &  27.7  & -3.05 &Fe II &1 Fe I .495\\
 272.754  & 1.50  & 7.12  & 8.60  &  30.4  & -3.05 &Fe II &  \\
 273.073  & 0.80  & 6.85  & 8.47  &  41.6  & -3.19 &Fe II &  \\
 273.245  & 1.45  & 7.10  & 8.46  &  22.8  & -4.92 &Fe II &1 Fe II .233\\
 273.697  & 1.50  & 7.12  & 8.60  &  30.3  & -3.22 &Fe II &1 Fe I .696\\
 273.731  & 2.00  & 7.25  & 8.27  &  10.5  & -5.12 &Fe  I &  \\
 273.955  & 1.43  & 7.10  & 8.61  &  32.1  & -2.31 &Fe II &  \\
 274.241  & 1.67  & 7.17  & 8.28  &  12.9  & -5.01 &Fe  I &1 Ti I .230, V II .243\\
 274.320  & 0.50  & 6.64  & 8.47  &  66.6  & -2.81 &Fe II &  \\
 274.407  & 2.50  & 7.34  & 8.27  &   8.4  & -5.51 &Fe  I &  \\
 274.648  & 0.90  & 6.90  & 8.47  &  37.0  & -2.59 &Fe II &  \\
 274.698  & 1.37  & 7.08  & 8.60  &  33.2  & -2.66 &Fe II &1 Fe I .698\\
 274.918  & 1.20  & 7.02  & 8.60  &  38.0  & -3.02 &Fe II &  \\
 274.932  & 1.07  & 6.97  & 8.46  &  30.9  & -2.38 &Fe II &  \\
 274.949  & 0.00  & 4.31  & 8.60 &\ldots  & -3.24 &Fe II &  \\
 275.014  & 1.58  & 7.14  & 8.25  &  12.8  & -5.04 &Fe  I &  \\
 275.574  & 1.17  & 7.01  & 8.46  &  28.1  & -2.17 &Fe II &  \\
 275.633  & 2.00  & 7.25  & 8.28  &  10.7  & -5.52 &Fe  I &2 Fe I .633, Cr II .630\\
 276.181  & 1.50  & 7.12  & 8.60  &  30.4  & -3.25 &Fe II &1 Fe I .178, Cr I .174\\
 276.893  & 1.50  & 7.12  & 8.60  &  30.3  & -3.08 &Fe II &  \\
 277.273  & 1.50  & 7.12  & 8.61  &  30.6  & -3.14 &Fe II &1 Fe I .283\\
  277.534  & 1.34  & 7.07  & 8.46  &  24.2  & -6.06 &Fe II &1 ?\\
  279.565  &  1.17  &  7.01 &   8.43    & 25.3 &    0.00 & Mg II  \\
 279.482    &1.98 &   7.24 &   8.57 &    21.3&    -5.65& Mn  I\\
 279.827&    1.70&    7.17 &   8.57 &    24.8 &   -5.79 &Mn  I  \\
 280.108    &0.84&    6.87&    8.57 &    50.3  &  -5.96 &Mn  I  \\
280.270  &  1.33& 7.07 &   8.43    & 21.9 &   -0.30 & Mg II \\
285.213 &   1.00    & 6.94 &   8.69 &    55.8 &   -2.20 & Mg  I \\ 
 393.366 &    1.17&    7.01 &   8.17&     14.3 &   -1.55 & Ca II  \\
 854.200 &    1.10 &   6.99 &   7.00&      1.0 &   -2.50 &Ca II  \\
\hline 
\end{tabular}\\
\normalsize
\end{center}
\vskip 4pt
Land\'e g-factors are computed using LS coupling. 
The final column lists the severity of the blend (as assessed by \citealp{Moore+Tousey+Brown1982}), and  the wavelength of
the blended line 
(decimal nm).  
$^\ast$The severity is: 0=unblended; 1=moderate blend; 2=severe blend. The data in Figure~\ref{fig:hanle}
are taken directly from this table. $^\dag$The strong line of \ion{Fe}{2} multiplet 1 
is blended with a line of multiplet 171.
\end{table*}
}

\shorttitle{}
\shortauthors{Judge et al.}

\begin{document}

\title{Measuring the magnetic origins of solar flares, CMEs and Space Weather}

\correspondingauthor{
Philip Judge}

\email{judge@ucar.edu}
\author{Philip Judge}

\newcommand{\hao}{
High Altitude Observatory,
National Center for Atmospheric Research,
Boulder CO 80307-3000,
 USA }
 
\affiliation{\hao} 
\author{Matthias Rempel}
\affiliation{\hao} 
\author{Rana Ezzeddine}
\affiliation{Department of Astronomy, University of Florida, 211 Bryant Space Sciences Center, Gainesville, Florida, 32611, USA} 
\author{Lucia Kleint}
\affiliation{Universit\'e de Gen\`eve,
Centre Universitaire d'Informatique,  
7, route de Drize,
1227 Carouge, Switzerland
}
\author{Ricky Egeland}
\affiliation{\hao} 
\author{Svetlana V. Berdyugina}
\affiliation{Leibniz-Institut fuer Sonnenphysik (KIS), Sch\"oneckstrasse 6, 79104, Freiburg, Germany}
\author{Thomas Berger}
\affiliation{Space Weather Technology, Research and Education Center,
3775 Discovery Drive
N433
Boulder, CO  80309}
\author{Paul Bryans}\affiliation{\hao}
\author{Joan Burkepile}\affiliation{\hao}
\author{Rebecca Centeno}\affiliation{\hao}
\author{Giuliana de Toma}\affiliation{\hao}
\author{Mausumi Dikpati}\affiliation{\hao}
\author{Yuhong Fan}\affiliation{\hao}
\author{Holly Gilbert}\affiliation{\hao}
\author{Daniela A. Lacatus}\affiliation{\hao}

\begin{abstract}
We take a broad look at 
the problem of identifying 
the magnetic solar causes  of space weather.   With the lackluster performance of 
extrapolations based upon 
magnetic field measurements in the photosphere, we identify a region in the near UV part of the spectrum as optimal for studying the development of magnetic free energy
over active regions. 
Using data from SORCE, Hubble Space Telescope, and SKYLAB, along with 
1D computations of
the near-UV (NUV) spectrum and 
numerical experiments based on the MURaM radiation-MHD and HanleRT radiative transfer 
codes, we address multiple challenges. These challenges are best met through 
a combination of near UV lines of bright  \ion{Mg}{2}, and lines of \ion{Fe}{2} and \ion{Fe}{1} (mostly within the $4s-4p$ transition array)  which form in
the chromosphere up to $2\times10^4$ K.  Both Hanle and Zeeman effects 
can in principle be used to derive 
vector magnetic fields.  However,
for any given spectral line 
the $\tau=1$ surfaces 
are generally geometrically corrugated owing to fine structure such as fibrils and spicules.  By using 
multiple spectral lines 
spanning different optical depths,
magnetic fields across nearly-horizontal surfaces can be 
inferred in regions of low plasma $\beta$, from which free energies, magnetic topology and other quantities can be derived.

Based upon the recently-reported successful suborbital space
measurements of magnetic fields with the CLASP2 instrument,
 we 
argue that a modest space-borne 
telescope will be able to make significant advances in the attempts to predict solar eruptions.  Difficulties associated with
blended lines are shown to be minor in an Appendix.
\end{abstract}


\section{Introduction}
\label{sec:introduction}

Commenting on the origin of solar flares, 
in 1960 Gold \&{} Hoyle wrote \nocite{Gold+Hoyle1960}
\begin{quote}
The requirements of the theory can therefore be stated quite definitely. Magnetic field configurations must be found that are capable of storing energy densities hundreds of times greater than occur in any other form, and that are stable most of the time. A situation that occurs only a small fraction of the time must be able to lead to instability in which this energy can rapidly be dissipated into heat and mass motion\ldots
\end{quote}
These requirements have remained unchanged in
the intervening six decades. We now have a ``standard model'' of flares 
\citep{Carmichael1964,Sturrock1966,Hirayama1974,Kopp+Pneuman1976},
involving  a loss of equilibrium of arcades of closed magnetic flux systems, particle acceleration and, frequently, ejection of plasmoids.  The instability can only arise when the free magnetic energy stored in the coronal magnetic fields exceeds certain thresholds. This is, however, a necessary but not sufficient condition for flaring and plasmoid ejection, because other global and local properties of the magnetic field, notably the topology, also determine the stabilility of coronal 
MHD systems \citep{Low1994}.

Enormous interest in solar eruptions 
is driven by society's need to 
understand origins of ``space weather''  
\citep[e.g.][]{Eddy2009}
as well as its intrinsic scientific challenges.  However, much like weather prediction before the advent of modern
instrumentation and computers, predicting the properties of eruptions and flares has remained a difficult challenge primarily due to the lack of quantitative data available for analysis of the pre-flare state and flare-triggering mechanism(s).  In particular, the only high-resolution, high cadence, data available to date has been the magnetic field state in the photosphere, which has been used to define ``boundary conditions" to the overlying coronal magnetic field, where most of the free energy is stored prior to 
eruption.  Herein lies the basic problem:
magnetic fields measured in 
the photosphere are not ideally suited to
the task. We elaborate upon this below.

The situation described by \cite{Gold+Hoyle1960} draws
analogy with trying to predict the timing and location of lightning strikes \citep{Judge2020}.  By measuring updrafts
and/or observing cloud buildup one can identify the causes of the buildup of electrostatic energy, but lightning occurs in
a predictable way only in a statistical
sense. In-situ measurements of electric fields are required even then to make 
probabilistic forecasts of lightning 
strikes to the ground, without reference
to intra- or inter- cloud discharges. 
In this sense the
weaknesses found in 
previous prediction methods is perhaps unsurprising. 

The difficulty of the task has been summarized in the write-up of an inter-agency ``All Clear Workshop'' \citep{2016ApJ...829...89B} in which the
authors conclude with the sobering statement that 
\begin{quote}
For M-class flares and above, the set of methods tends toward a weakly positive skill score (as measured with several distinct metrics), with no participating method proving substantially better than climatological forecasts.   
\end{quote}

Steady improvements in the  performance of flare forecasts has been documented in more recent years \citep[e.g.][]{Lekaetal2019}.  However, these authors make explicit the need to augment existing photospheric magnetic data in order to proceed, with a heterogeneous mix of theory and observation:
\begin{quote}
    quantitative ``modern'' forecasts incorporate \ldots physical understanding as they often characterize coronal magnetic energy storage by proxy, with the parameterizations of photospheric magnetograms. 
\end{quote}
In other words, 
measurements of the photospheric magnetic field can be related only \textit{indirectly} and through certain \textit{ad-hoc} 
parameterizations, to the magnetic free energy of the overlying corona, which is the ultimate origin of
instabilities leading to flares and CMEs.

With the availability of a decade  of high-resolution photospheric magnetic field data from NASA's Solar Dynamics Observatory (SDO) mission, recent efforts have focused on statistical pattern recognition methods for flare prediction that generally fall under the ``machine learning (ML)'' sub-field of artificial intelligence research. For example, a study by \cite{bobracouvidat2015} used photospheric vector magnetic data of 2071 active regions from
the Helioseismic and Magnetic Imager (HMI) instrument on SDO. They predicted strong flares using a Support Vector Machine (SVM) algorithm applied to a set of 25 parameters derived from the vector magnetic field measured in SHARPs (Space-weather HMI Active Region Patches), which are cut-outs around sunspot active regions.
They achieved a high True Skill Statistic (TSS) of $\approx$ 0.8, but at the cost of getting a significant number of false positives, a common characteristic of all ML flare prediction algorithms optimized for TSS in the strongly unbalanced training sets which naturally contain many more ``no flare'' examples than ``flare'' examples. 
Later studies \citep{Park+others2018,Liu+others2019,Chen+others2019,Li+others2020} 
applied recurrent architectures such as Long Short-Term Memory (LSTM) systems and deep learning systems including Convolution Neural Network (CNN) and autoencoder architectures to explore feature sets that extend beyond the magnetic field-derived features of earlier ML flare prediction systems. However these studies do not demonstrate significant increases in skill scores relative to earlier approaches. 

These studies have used only photospheric measurements of
magnetic fields. Investigations have also been carried out that include information from higher atmospheric layers, e.g., by incorporating SDO Atmospheric Imaging Array (AIA) data into existing ML flare prediction systems \citep{nishizuka+others2018,jonasetal2018},  
or by using extrapolated coronal magnetic field models \citep{Korsos+others2020}. 
Chromospheric UV spectral lines from IRIS have also been 
scrutinized for additional flare prediction utility \citep{panoskleint2020}.  
These studies all show some degree of improvement in flare prediction metrics when including chromospheric or derived coronal information. 
More recently, \cite{Deshmukh+others2020} showed that combining Topological Data Analysis (TDA) of the radial magnetic field structure in active regions with the SHARPs vector magnetic field metadata results in similar or slightly higher skill scores compared to using selected SHARPs metadata alone. This result implies that inclusion of high-resolution imaging of active region and flows may be an important complement to spectroscopic measurements when deducing atmospheric conditions that evolve to a flaring state. 

Physically, however, solar flare prediction 
based upon photospheric data is ultimately confronted with 
basic difficulties, beyond the well-known
disambiguation problem associated with native 
symmetries in the Zeeman effect.   The photosphere is a dense radiating layer of
high density ($10^{-7}$ g~cm$^{-3}$) plasma.
Outside of sunspot umbrae the photospheric magnetic field
is unable to suppress convective motions which
therefore control the magnetic fields
threading the fluid, a case where the 
plasma-$\beta=$ gas pressure /
magnetic pressure, is generally $>1$.  
In traversing the chromosphere, emerging magnetic fields drop in strength algebraically %geometrically?
but the plasma densities and energy densities 
drop exponentially (by 7 and 5 orders of magnitude, respectively), so that the tenuous plasma
at the coronal base is in a low-$\beta$ state. 
One immediate practical
problem is that tiny dynamical changes which 
may be unobservable in the dense 
photosphere can have enormous consequences for the tenuous plasma and magnetic fields above: the problem is observationally ill-posed.  
 As a result of
these different photospheric and coronal plasma regimes, 
the time scales for significant photospheric magnetic field evolution are typically measured in minutes to hours whereas the dynamic evolution and reconnection of the coronal magnetic field configurations leading to flares,
after a quasi-static buildup phase, is measured in seconds. This fundamental mismatch in dynamical time scales evidently cannot be overcome through any type of analysis of photospheric magnetic field data and leads to the conclusion that quantitative analysis of the magnetic field and flows in the chromosphere and corona are necessary to catalyze significant breakthroughs. 
%in our ability to predict solar magnetic eruptions and subsequent space weather impacts at Earth. 

%%%%%%%%%%%%%%%%%%%%%%%%%%%%%%%%%%%%%%%%%%%%%%%%%%%%%%%%%%%%%%%%%%%%%%%
\begin{figure*}[ht]
\includegraphics[width=0.32\linewidth]{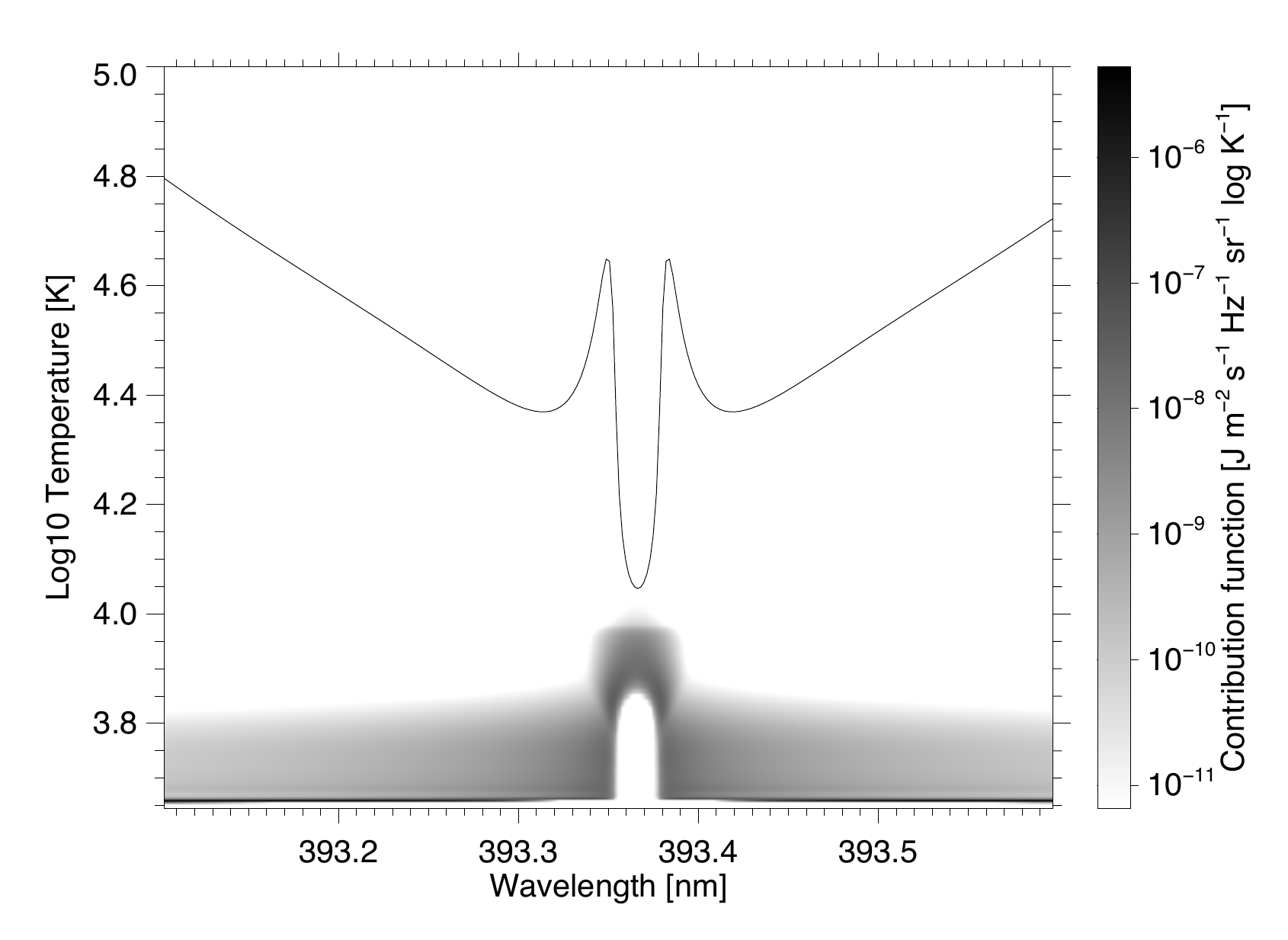}
\includegraphics[width=0.32\linewidth]{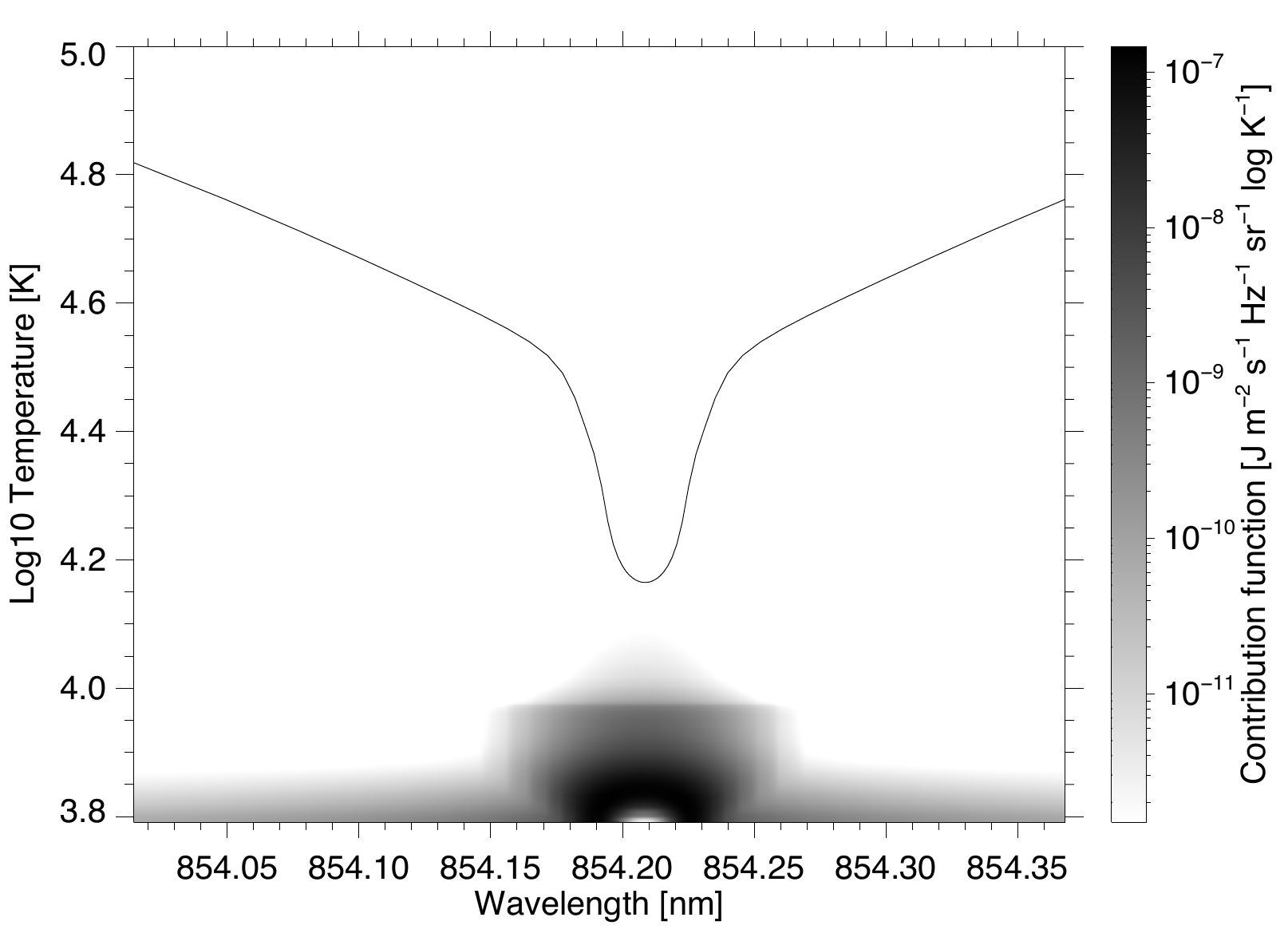}
\includegraphics[width=0.32\linewidth]{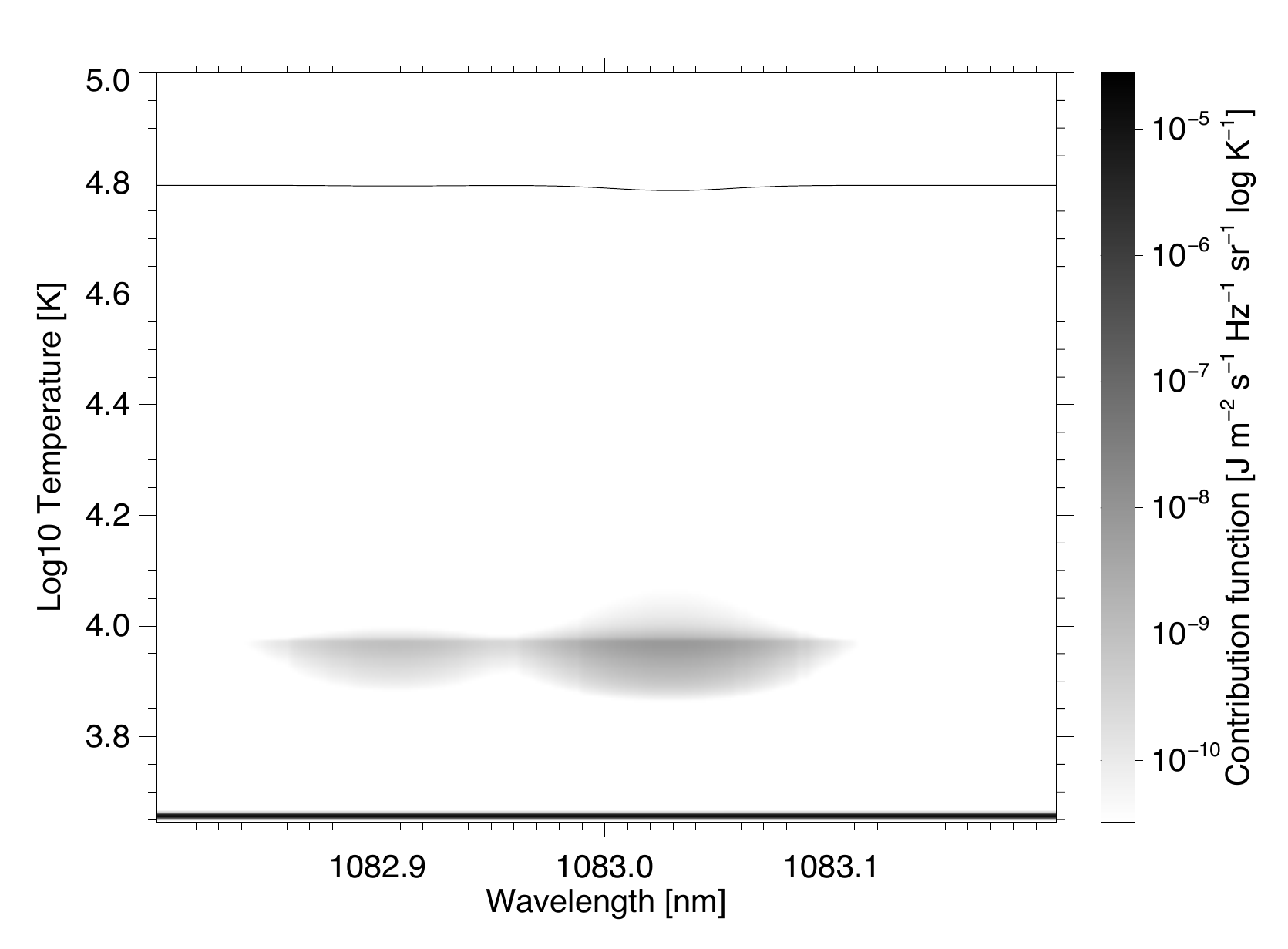}
\\
\includegraphics[width=0.32\linewidth]{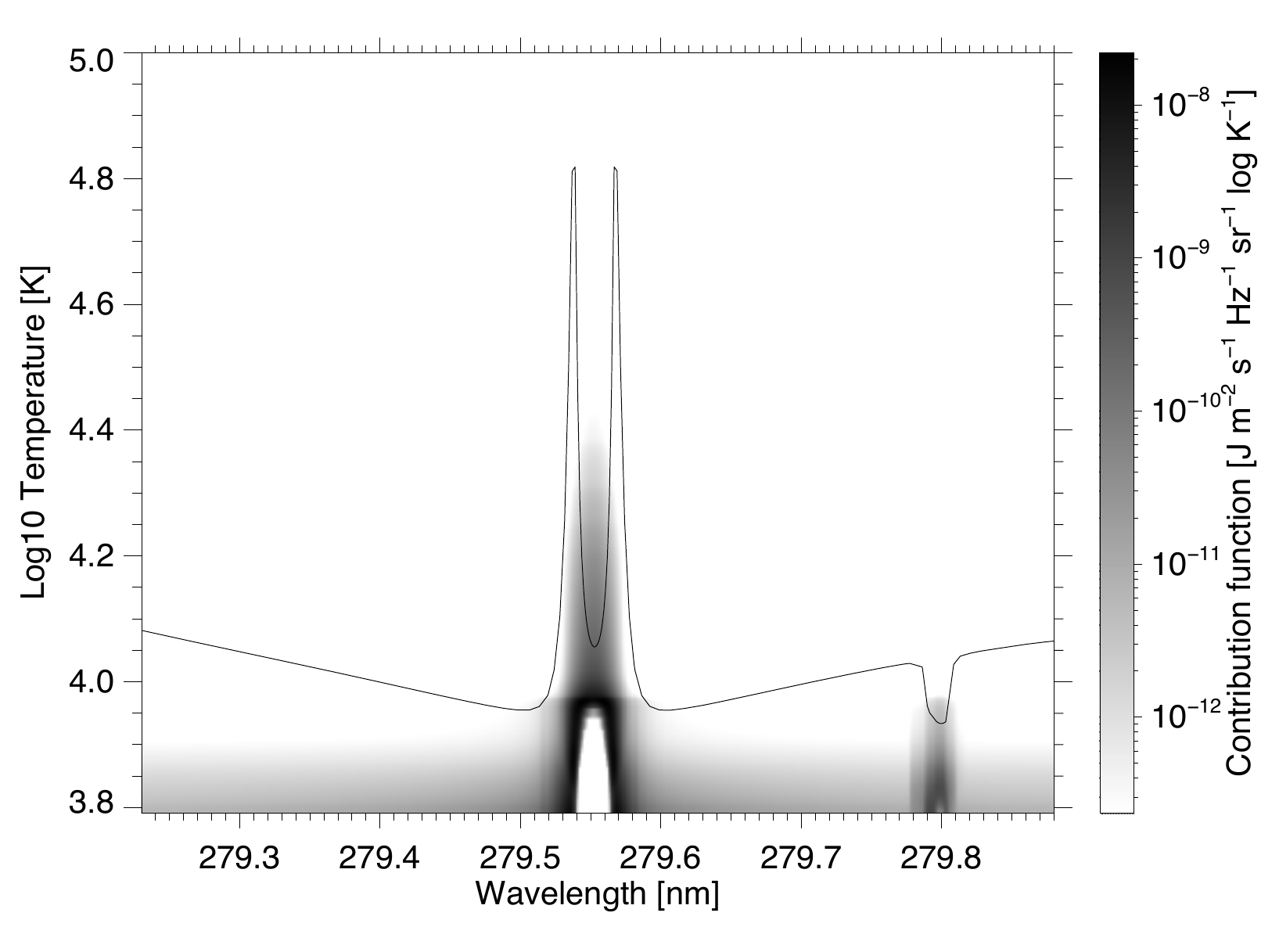}
\includegraphics[width=0.32\linewidth]{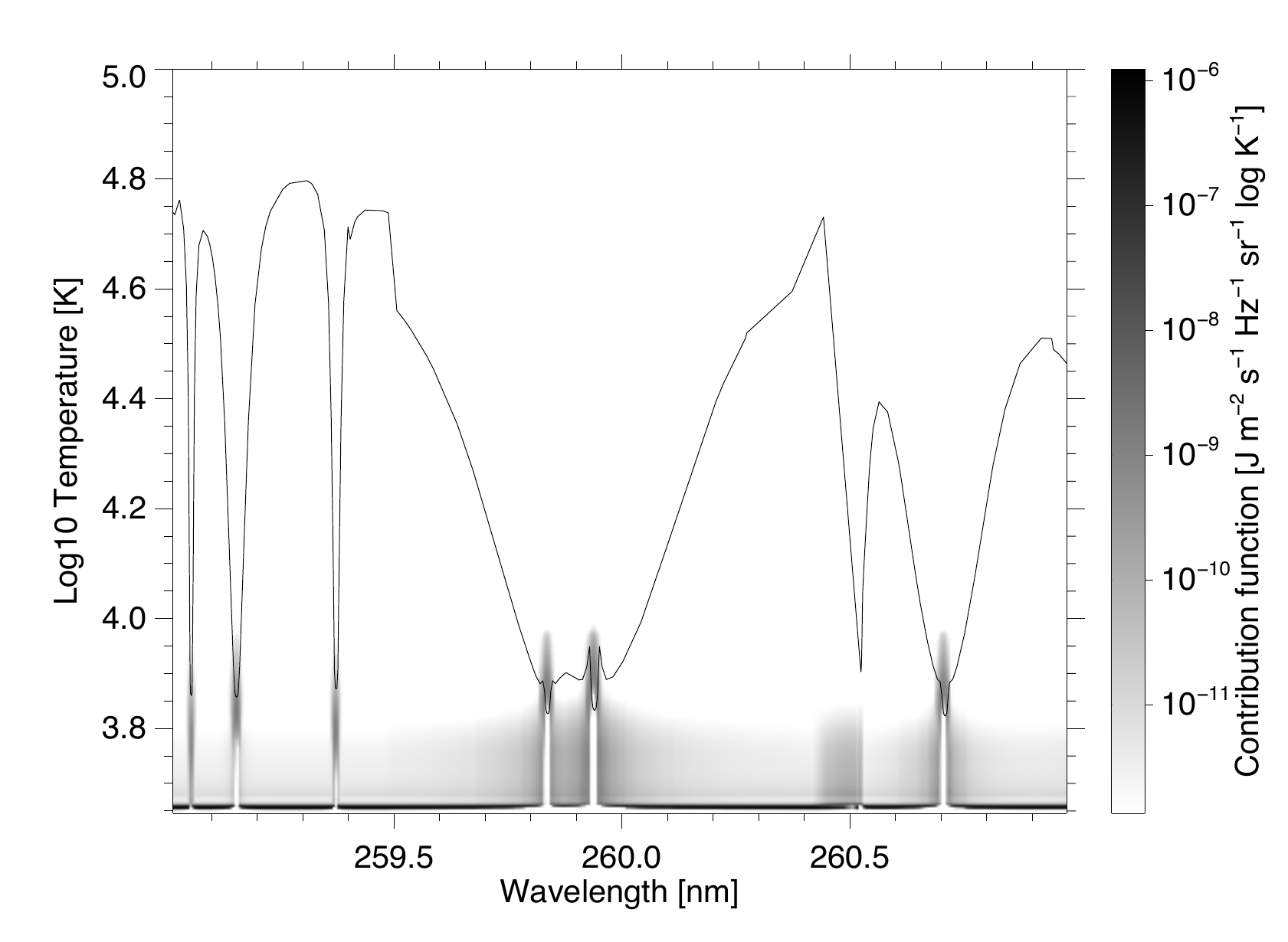}
\includegraphics[width=0.32\linewidth]{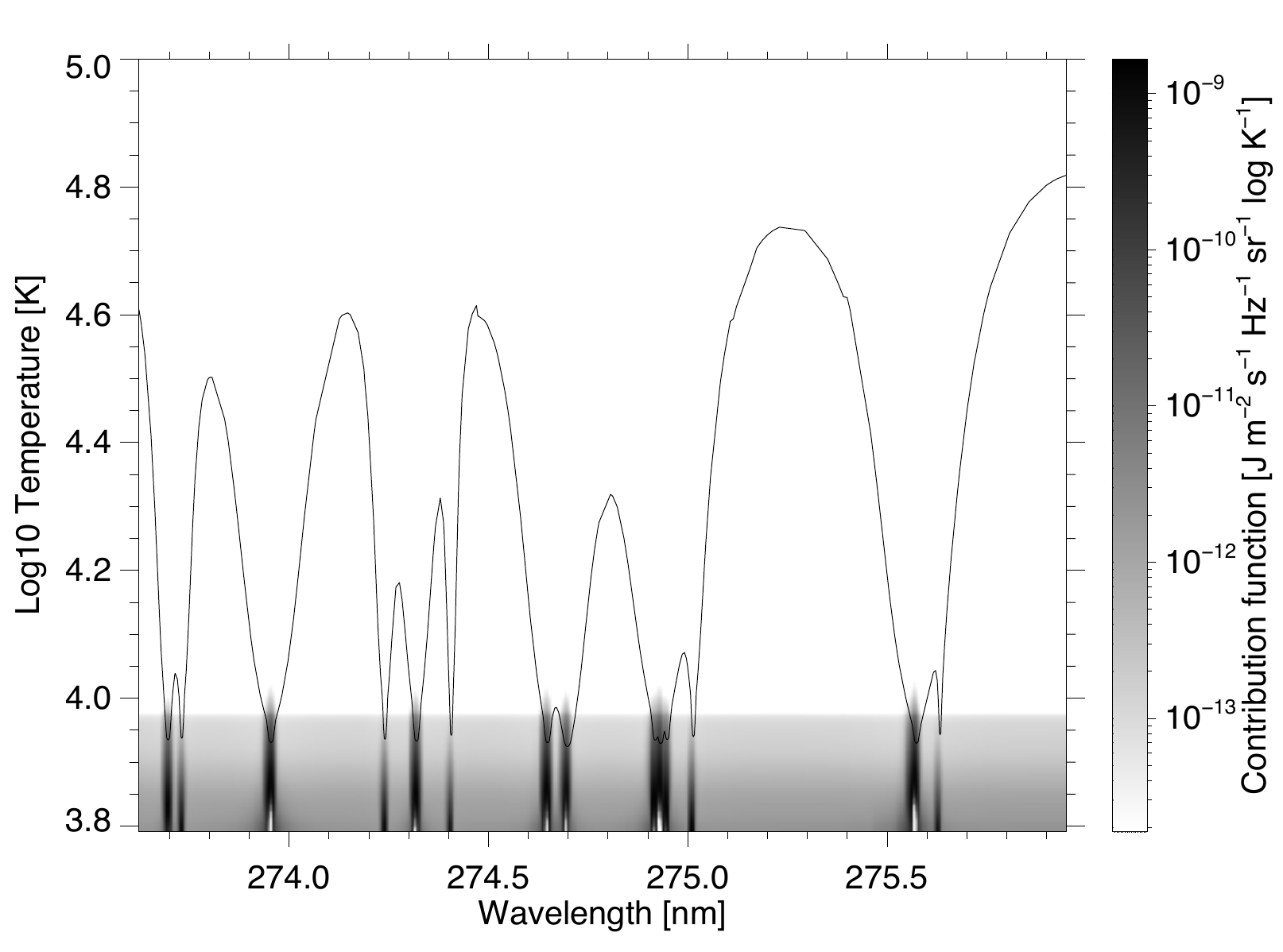}
\caption{Contribution functions are shown 
as a function of temperature from the
model C of 
\cite{Fontenla+Avrett+Loeser1993}.  The emergent intensities 
are over-plotted
as solid lines for 
a ray from close
to disk center ($\cos \theta = 0.887$, $\theta$ the angular distance from disk center as measured from Sun center). 
The \ion{Ca}{2} lines at 393 and 854 nm span from 3.8 to 
4.0 in log$T$ (the latter 
forming in deeper cooler layers 
than the former), 
and the 1083 nm
line of \ion{He}{1} forms within the last pressure scale height of
the chromosphere.   
Notice the extension 
of contributions for the \ion{Mg}{2} resonance lines beyond all others, to $\log T_e \ge 4.3$, and the multiple lines of \ion{Fe}{2} between 259 and 261 nm whose cores form close to 
3.9-4.0 in log$T$, and weaker lines in the 275 nm region
with stronger contributions from lower 
temperatures.  A modest atomic model for \ion{Fe}{2} was adopted 
to examine the line cores formed  in non LTE in the upper chromosphere.  The calculations used the RH computer program \citet{Uitenbroek2001}. 
Far more complete LTE
calculations are made in the Appendix where spectral line blends are 
discussed. 
}
\label{fig:contrib}
\end{figure*}

In order to address this and other questions, our purpose here is to
\textit{explore the entire solar spectrum to attempt to identify more direct ways to identify the free energy and
topology of magnetic fields above active regions.}
In this way we hope 
to put
the prediction of solar magnetic eruptions, the source of all major space weather events, on a firm observational basis. The specific problem we address is reviewed 
in section
\ref{sec:overview}. It will involve 
spectro-polarimetry.

Before 
proceeding, we point out that 
our goals may on first reading appear 
similar 
to efforts 
measuring only 
intensities of EUV and UV 
emission lines \citep[e.g.][]{dePontieu+others2020}, an endeavor with a history of 6+ decades yet 
able to study only the effects and not causes of magnetic energy storage and release. Instead our work is closer 
to an important study
by \citet{Trujillo+others2017}, but the latter focuses more on the physical processes behind the magnetic imprints across the solar UV spectrum, including the near-UV spectral region of the \ion{Mg}{2} $h$ \&{} $k$ lines. We focus instead on 
finding an optimal set of observations 
which target the problem of the build-up and release of magnetic energy.   We argue that spectro-polarimetry of the near-UV solar spectrum is indeed a 
profitable avenue for research.

\section{Observational signatures of energy build-up and release}
\label{sec:overview}

\subsection{Measuring magnetic fields in and above chromospheric plasma}

Two approaches are in principle
possible: \textit{in-situ} measurements and remote sensing. 
Clearly \textit{in-situ} methods  are far beyond the capabilities of known technology.
The state-of-the-art mission Parker Solar Probe \citep{Fox+others2016} 
has a perihelion 
distance of just  below ten solar radii, far above active regions
where most of the free magnetic energy exists. Therefore we limit our discussion to measuring magnetic fields remotely, an area which has a firm
physical basis \citep{Landi+Landolfi2004}.  

Briefly, two types of remote-sensing measurements have been made: direct measurements of 
light emitted from 
within coronal plasma, and measurements close to
the corona's lower boundary,
some 9 pressure scale heights 
above the photosphere.  Coronal
measurements include
\citep[see, for example, reviews by][for more complete references]{Judge+others2001,Casini+White+Judge2017,Kleint+Gandorfer2017}:
\begin{itemize}
    \item Measurements of line-of-sight (LOS)  field strengths using differences between ordinary and extraordinary rays of magneto-ionic theory at GHz frequencies \cite{Gelfreikh1994}.
    \item Measurements of field strengths across surfaces determined typically by a  harmonic of gyro-resonant motions of electrons at frequencies of several GHz,  frequencies uncontaminated by bremsstrahlung 
    opacity only when the field strength exceeds $\approx$200 G \citep{White2004}.
    \item Serendipitous level-crossings in atomic ions produces changes in EUV spectra through level mixing by the Zeeman effect 
    \citep{Li+others2017}, yielding magnetic field strengths.
    \item A long history of measurements of magnetic dipole lines during eclipses or with coronagraphs hold great promise for measuring vector magnetic fields in the corona.  The infrared, coronagraphic DKIST facility offers significant new opportunities for advancements using such lines  \citep[e.g.][]{Judge+Habbal+Landi2013}.
    \item Occultations of 
    background radio sources by the corona have revealed  integrated properties of 
    coronal magnetic fields, 
    weighted by electron density 
    through the Faraday effect
    \citep{Bird+Edenhofer1990}, a technique also applied to  signals from spacecraft away from the Sun-Earth line \citep{Bird1982}.

\end{itemize}
Of the second type, we can list:
\begin{itemize}
    \item Several strong lines measured using spectro-polarimetric measurements of
    the Zeeman effect from the ground can yield LOS components of chromospheric magnetic fields.  Very high photometric precision is however required for vector fields, beyond most current  capabilities  \citep{Uitenbroek2010}, owing 
    to the second-order $\epsilon^2$ signature of the Zeeman effect in the small ratio $\epsilon$ of magnetic splitting to the line Doppler width (defined below). 
    These include 
    \ion{Na}{1} D lines, \ion{Ca}{2} \textit{H, K} and infrared triplet lines.  Synthetic \ion{Ca}{2}  line profiles are shown in
    the non-LTE calculations using     
the RH non-LTE  program including partial redistribution  \citep{Uitenbroek2001}  
    in 
    Figure~\ref{fig:contrib}.

    \item The  \ion{He}{1} 1083 nm triplet has a proven record of magnetic field determinations 
    in a variety of cool structures such as spicules \citep{Centeno+others2010},
    filaments \citep{2012A&A...539A.131K,     
    2019A&A...625A.128D} 
    and 
    cool loops \citep{Solanki+others2003}, in the vicinity of the base of the corona before, during and after flares     
    \citep{Judge+others2014}
    even extending far into the corona  \citep{2012AAS...22052119J,    
    Schad+others}.   Both the Zeeman and Hanle effects contribute to the formation of this line, owing to the large contribution of anisotropic radiation incident on the helium atoms to the statistical equilibrium in  the triplet states  \citep[e.g.][]{2007ApJ...655..642T}.
    \item Recently, sub-orbital rockets have carried sensitive spectropolarimeters to observe the resonance lines of hydrogen \citep{2017ApJ...839L..10K}    
    and singly-ionized magnesium \citep{Trujillo2018,Ishikawa+others2021}.    
    Both Zeeman and Hanle effects have revealed information on magnetic fields near the very top of the solar chromosphere. 
    \item Measurements at 30 GHz and above made with the ALMA interferometer facility can in principle yield  line-of-sight 
     magnetic field strengths  using  Gelfreikh's (1994)  theory \new{of continuum polarization \citep{Loukitcheva2014, Loukitcheva2020}, but we 
     have been unable to find reports of solar  polarization measurements with ALMA. It} can observe from 0.4 to 8.8 mm wavelengths (750 and 34 Ghz respectively), thereby  sampling  
      heights from 600 km to 2000 km in the chromosphere. \new{Transition region plasma  under non-flaring conditions is tenuous and thin. It contributes very little to the far brighter chromospheric emission at the high frequencies measured by ALMA. }
    
\end{itemize}
 Next we assess the ability of
these 
techniques to address the specific problem
of interest.

\begin{figure*}[ht]
\includegraphics[width=\linewidth]{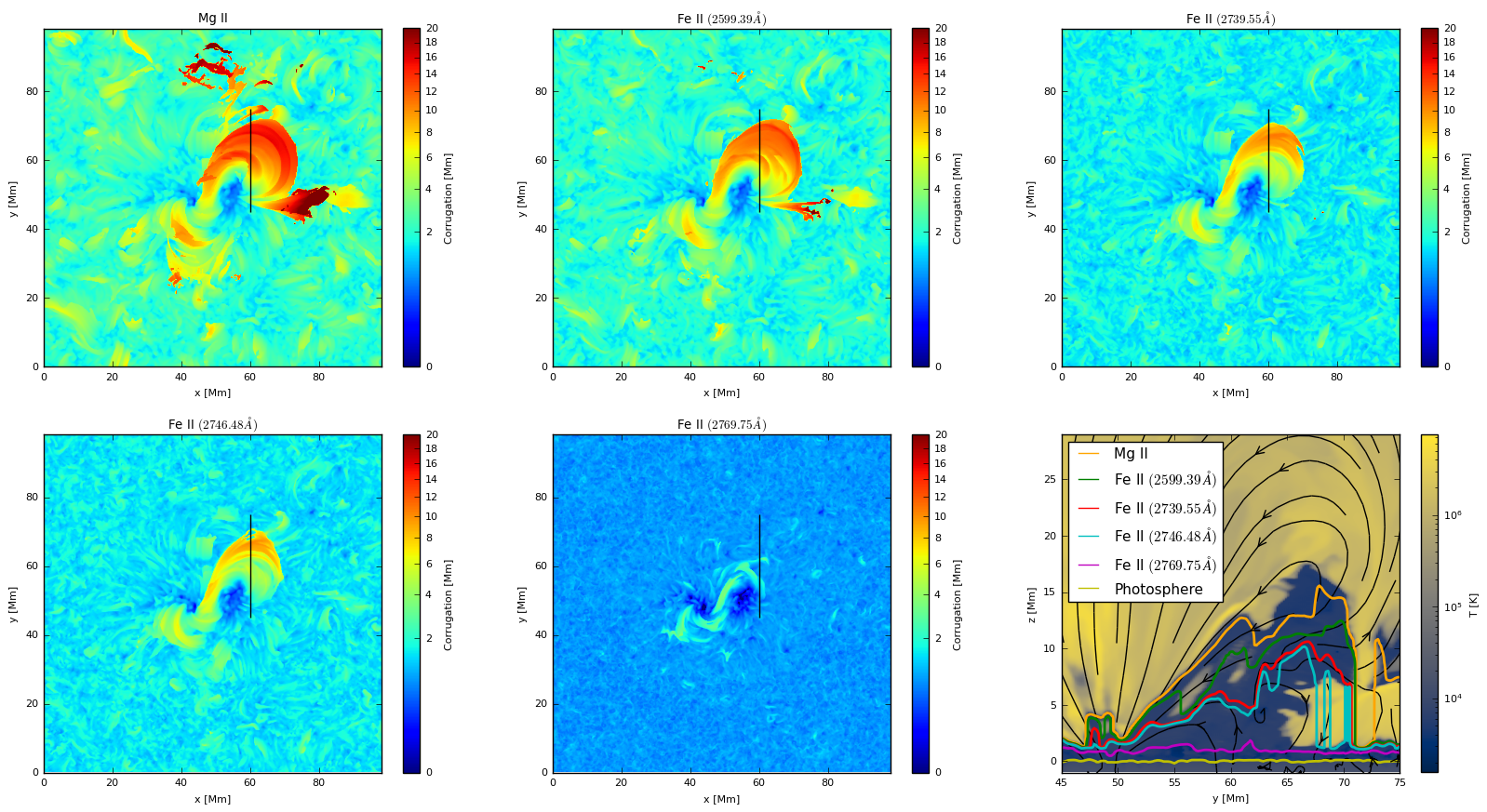}
\caption{Images are shown to highlight the corrugated 
surfaces at which the centers of UV lines are formed (close to $\tau=1$) in MURaM 
calculations of the upper solar chromosphere.  The calculations are from a numerical experiment where 
a magnetic flux system is emerging from
beneath the solar atmosphere. The emergent fields carry with them 
plasma, which is revealed by 
the extended heights of the $\tau=1$ surfaces, which can 
exceed 10 Mm. 
Typically the line shifts in the lower transition region
are smaller than the line widths \citep{Athay+Dere1991}. 
Therefore the $\tau=1$ surfaces were  
computed without taking into account Doppler shifts,
at the centers of lines of \ion{Mg}{2} $k$, and several lines of \ion{Fe}{2}. The $x-y$ images of height 
of formation are plotted in order of decreasing 
opacity from \ion{Mg}{2} to \ion{Fe}{2} 2769.75 \AA. 
In the bottom right panel, slices of these $\tau=1$ surfaces 
are shown, taken along the black line in the five 
other panels.  But the black lines in this 
final panel are magnetic lines of force in the $y-z$ plane. Evidently, a combination of these and other lines of iron with a wide range of opacities 
spans the ranges of
excursions of the surfaces away from a horizontal plane. }
\label{fig:corrugated}
\end{figure*}

\subsection{Establishing methods to measure free magnetic energy and topology}

While information on
magnetism above the photosphere is contained in all the above strategies, few 
offer ways to assess the necessary quantities concerning magnetic free energy as magnetic fields emerge, build and change topology as it evolves and is suddenly released. If we consider measurements within the corona, we would have to probe the 3D magnetic structure. However, the corona is optically
thin (with the exception 
of gyro-resonance emission which yields just the magnetic field strength within resonant surfaces, whose geometry is not known from 
one line of sight), 
which means that recovery
of the 3D coronal field  would require 
measurements from at least
two different lines of sight
\citep[e.g.][]{Kramar+others2006}.
Until suitable  instruments are flown on spacecraft away from the Sun-Earth line, the required stereoscopic views will be unavailable.
Therefore we must seek 
alternatives. 

When chromospheric, transition region and coronal 
plasmas are connected by magnetic lines of force 
\citep{Judge2021}, 
the plasma pressure near the 
top of the chromosphere 
is close to that of the overlying corona. Accurate measurements of
magnetic fields using spectral
lines formed in plasma up to a few times $10^4$ K can therefore
be used to inform us on 
the evolving magnetic state 
of the overlying corona \citep{Trujillo+others2017}.   
Selected spectral line profiles formed
across the upper chromosphere 
are shown in Figure~\ref{fig:contrib}, intended to illustrate 
near UV lines in comparison to familiar lines commonly observed 
from the ground.   The solar spectrum between 250 and 281 
nm is dominated by line transitions of \ion{Mg}{2} and 
many lines belonging to
the $4s-4p$  transition array in neutral and 
singly ionized iron. 
In the figure, contributions to the intensities of \ion{Mg}{2} and typical \ion{Fe}{2} lines
are  shown as a function of
electron temperature. This abscissa was chosen instead of
height and continuum optical depth because in the particular
1-dimensional model adopted, the temperature
gradient in the transition region is very large, making lines appear to originate from the 
same heights.  

The regime of plasma $\beta=8\pi p/B^2 < 1$ in upper chromospheric, transition region and coronal lines is the same, if the photospheric 
magnetic field has expanded to
fill the volume.  In active region plasma this appears
to be the case \citep{Ishikawa+others2021}.
When measurements are 
made close to the force-free state ($\beta \ll 1$), 
then theory can be 
reliably invoked in at least two ways: application of the magnetic virial theorem \citep{Chandrasekhar1961}, and extrapolation of fields with observational boundary conditions 
compatible with the 
magnetostatic conditions \citep{Wiegelman+Sakurai2021}. Any waves and/or tangential discontinuities 
measured will also be of importance to assessing the free magnetic energy and its evolution.  \cite{Fleischman+others2017} have confirmed 
these theoretical concepts quantitatively using 
model simulations.

Skeptics of our 
statements 
might rightfully
 recall that the chromosphere is finely structured
 \citep[see, e.g., the remarkable structure in  narrow-band images of][]{Cauzzi+others2008,Wang+others2016,Robustini+others2018}.  It is also known, for example from limb observations, that transition region plasma at intermediate temperatures has contributions from structures far from a level surface which might otherwise be amenable to mathematical techniques to estimate free energies and extrapolated magnetic fields.  A 
 non-level, ``corrugated'' surface of unit optical depth
 in a given spectral line 
can lead to spurious inferences 
 of magnetic structure 
 even within low-$\beta$ plasmas \cite[see, e.g., section 4.3 of the paper by][and our 
 Figure~\ref{fig:corrugated}
below] 
 {Trujillo+others2017}.
  The question then arises, \textit{is the interface between 
 chromosphere and corona just too finely structured to make measurements of magnetic fields there of little use?} 
 
 Hints from two pieces of work suggest that this is not the case.  Firstly, the CLASP2 measurements reported by 
 \cite{Ishikawa+others2021} have shown that, at an angular resolution of 1" which is far smaller than the scales of large coronal structures leading to flares, the fields measured in the core of the \ion{Mg}{2} $h$ \&{} $k$ lines across active regions are unipolar and spatially smooth over large areas.  Any mixed polarity fields
 undetected by CLASP2 must 
 lie below 1" scales and will 
 therefore not penetrate more than a few hundred km above the 
 photosphere.  But the fields 
 measured by CLASP2 
 may well contain unseen tangential
 discontinuities arising from
 small-scale angular displacements across flux surfaces \citep{Judge+others2011}. Such 
 features would constitute 
 an additional source of free magnetic energy whose consequences might then be explored using the very techniques advocated for below.
  Secondly, 
 \cite{Judge2021} 
has shown that the 
attention to the dynamics and fine structure of 
transition region plasma has been 
over-emphasized, statistically.  Most of the chromosphere-corona transition region on the Sun is indeed in a thermal interface connected by magnetic lines of force, and not isolated, as was 
claimed by \cite{2014Sci...346E.315H}
and references therein, all of them based
on circumstantial 
evidence accumulated using
spectral line intensities alone. 

One example is shown in Figure~\ref{fig:corrugated}. This figure is based on a MURaM flux emergence simulation in a 98 Mm wide domain, reaching from -8 to 75 Mm in the vertical direction. The snapshot corresponds to a highly energized corona, in which a strong bipolar field has emerged into a previously quiet region.  Ten minutes after this, a flare and  a mass ejection occurred. The simulation is based on the \citet{2017ApJ...834...10R} version of MURaM, which uses a simplified description of the chromosphere. The $\tau=1$ surfaces were computed using 
parameterized li<ne opacities at line center
and integrating downwards from
the corona.   The opacities 
were derived using a function, which for the \ion{Mg}{2} $k$-line for example, are of the form 
\begin{equation}
    \kappa_{\nu=\nu_0} 
    = \frac{\pi e^2 f_{ij}}{m_e c} 
    \frac{n_{\mathrm{i}}}{n_{\mathrm{Mg^+}}}     \frac{n_{\mathrm{Mg}^+}}{n_{\mathrm{Mg}}} \frac{n_\mathrm{Mg}}{n_\mathrm{H}} n_{\mathrm H} \phi_{\nu_0},    
\end{equation}
in which the lower level population $n_\mathrm{i}$ is expanded as a product of ionization
fraction $ \frac{n_{\mathrm{Mg}^+}}{n_{\mathrm{Mg}}}$ (a function of
electron temperature $T_e$), 
the abundance 
$\frac{n_\mathrm{Mg}}{n_\mathrm{H}}$, and the hydrogen density $n_{\mathrm H}$ from the model.  The first factor is the frequency-integrated 
cross section of the $k$ line 
from fundamental constants and
the absorption oscillator
strength $f_{ij}$ between levels $i$ and $j$,  
and
$\phi_\nu$ the line profile at frequency $\nu$, normalized so that $\int_{0}^\infty \phi_\nu d\nu=1$. The profile is 
dominated for these ions by
an assumed microturbulence of 10 km~s$^{-1}$.
The term $ \frac{n_{\mathrm{Mg}^+}}{n_{\mathrm{Mg}}}$ 
is computed assuming that neutral Mg is negligible, but allowing for electron impact ionization 
to Mg$^{2+}$ and higher ions and radiative recombination 
using data from \cite{Allen1973}.

This figure illustrates  approximate line formation heights, sufficient for the 
present paper.  The excursions in height corresponding to the emerging flux system vary from a
maximum near 10 Mm for \ion{Mg}{2} $k$, to a few Mm
for a line of \ion{Fe}{2} at 276.9 nm.
As can be seen, these corrugations can be probed with multiple spectral
features with different opacities.  Thus measurements at the top of the chromosphere, 
properly sampling the potential excursions of the 
isotherms and isobars in the transition between 
chromosphere and corona, appear to be a promising approach to the problem.  At the very least, 
such new magnetic measurements 
could be compared in detail with advanced  simulations of the Sun's atmosphere \citep{Gudiksen+others2011,2017ApJ...834...10R}.  Below we will suggest 
one approach to this problem based upon machine learning techniques.

\begin{figure}[ht]
\includegraphics[width=\linewidth]{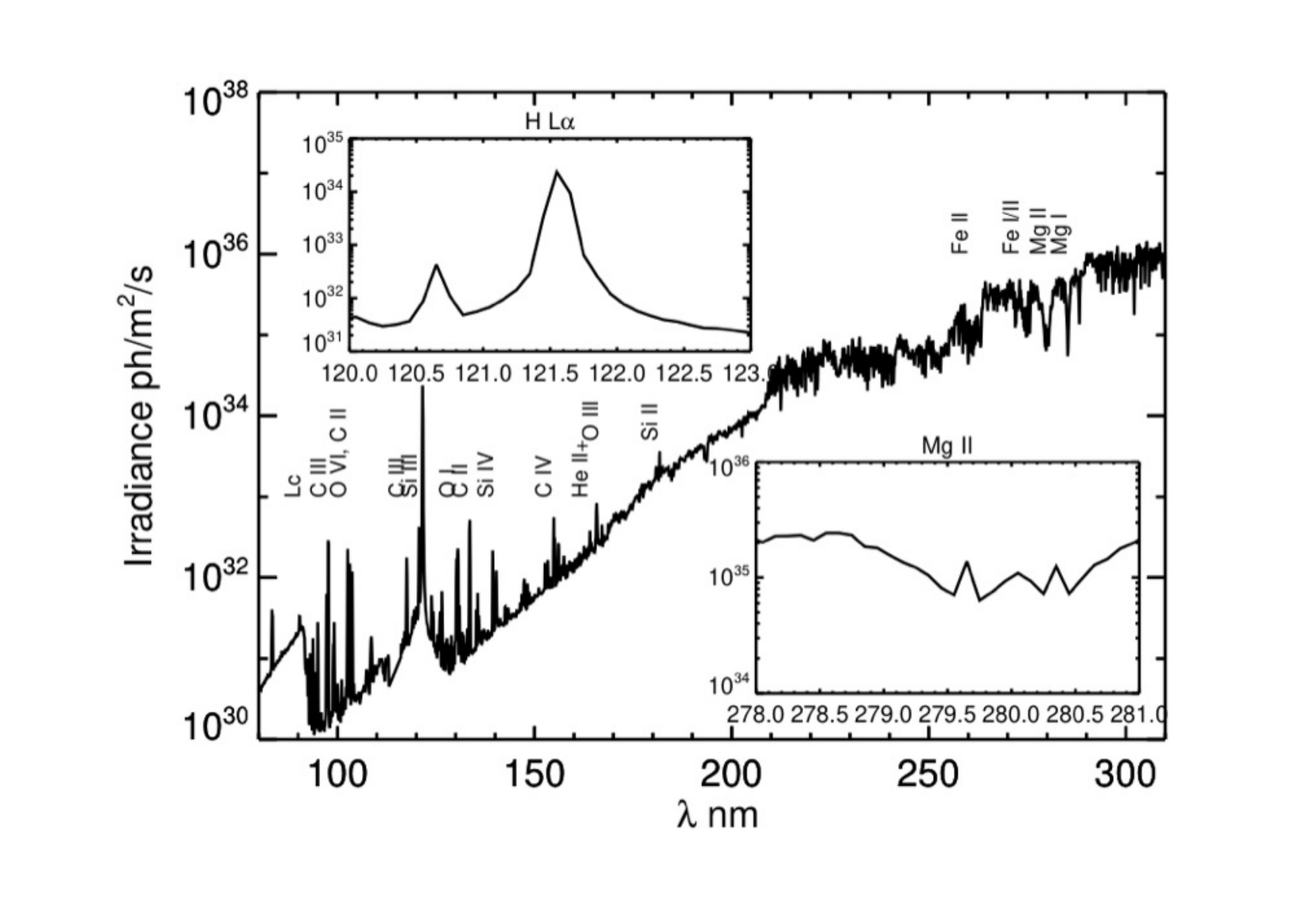}
\caption{A low-resolution UV irradiance spectrum from the Whole Heliospheric Interval in 2008 is shown (from {
https://lasp.colorado.edu/lisird/data/whi\_ref\_spectra}).   The near UV region 
has a photon flux density far higher than the majority of vacuum UV lines below 200 nm.   The exception is H L$\alpha$ which roughly 10x smaller than the flux density at 260 nm and longer wavelengths.
    }
\label{fig:alluv}
\end{figure}

\subsection{The near UV solar spectrum and the chromosphere-corona interface}

A standard, low resolution solar UV spectrum is shown in Figure~\ref{fig:alluv}.
After an extensive search, we have identified 
a spectral range 
of 256-281 nm as 
a promising part of the solar spectrum 
for measurement of vector  magnetic fields 
at the base of the corona. 
Our search, driven 
by the need to probe in detail magnetic structure 
in height above the solar surface \new{and/or depth
along the LOS of prominences, }
led to the choice of 
important transitions 
in the 4s-4p transition 
array of \ion{Fe}{2}.  While
there are strong \ion{Fe}{2} 4s-4p transitions  
in the region near 234 nm, 
the solar fluxes are weaker than in the 256-281 nm range. \new{Our wavelength selection
is significantly broader than the 279.2-280.7 nm spectral region of
CLASP2, which includes the Mg II $h$ \& $k$ line cores and 
magnetically-sensitive wings  
\citep{aueretal1980,2012ApJ...750L..11B, Alsina+others2016, delPinoAleman:2016}
the 3p-3d
lines \citep{delPinoAleman:2020}, and
photospheric lines of \ion{Mn}{1} 
\citep{Ishikawa+others2021}. }
%The wider wavelength range is needed to 
%sample multiple surfaces of unit optical depths required to mesure magnetic structure and topology in 3D. }
Both Hanle and Zeeman effects 
can be brought to bear on
the problem depending on
the strength of the magnetic fields found across the solar atmosphere. The Hanle effect can be important to remove the 
180$^o$ ambiguity present in Zeeman effect measurements.  
The rationale for this spectral
region is based upon  requirements demanded by the desire to measure vector magnetic fields within force-free ($\approx$ low $\beta$) plasma
at the coronal base, over an area
covering a typical active region. An angular resolution sufficient to resolve the thermodynamic 
fine structure 
is also a
constraint. Such features are 
most clearly seen only in 
images of features within 
the narrow spectral lines 
chromospheric images.  This constraint poses difficulties 
for current observations at radio
and sub-mm wavelengths, and in hard X-rays.  
Soft X-rays ($\le 1$ keV) have no known sensitivity to  magnetic fields of
solar magnitude, and along with EUV photons cannot escape from or penetrate deep into
the chromosphere. 
We are then left to consider 
vacuum UV to infrared wavelengths.  

We further eliminate wavelengths above the atmospheric cutoff at 310 nm 
because of the lack of spectral lines that can 
span many decades in optical depths above the chromosphere.  The Balmer and higher line series of 
hydrogen are poor choices
for Zeeman diagnostic work,
the lightness of hydrogen 
makes lines broad and 
patterns overlapping. 
For Ba-$\alpha$, seven anomalous Zeeman patterns will lie within half a wave number for typical solar field strengths even in the photosphere, thereby making the observation of anomalous Zeeman patterns extremely difficult.  The Hanle effect has been explored by
\cite{2010ApJ...711L.133S} in their comparison of 
\new{1D radiative transfer calculations
with limb observations of the quiet Sun by 
\citet{Gandorfer2000}. The 
value of H$\alpha$ as a diagnostic of magnetic field remains a subject for
further research.}

\begin{figure}[ht]
\includegraphics[width=\linewidth]{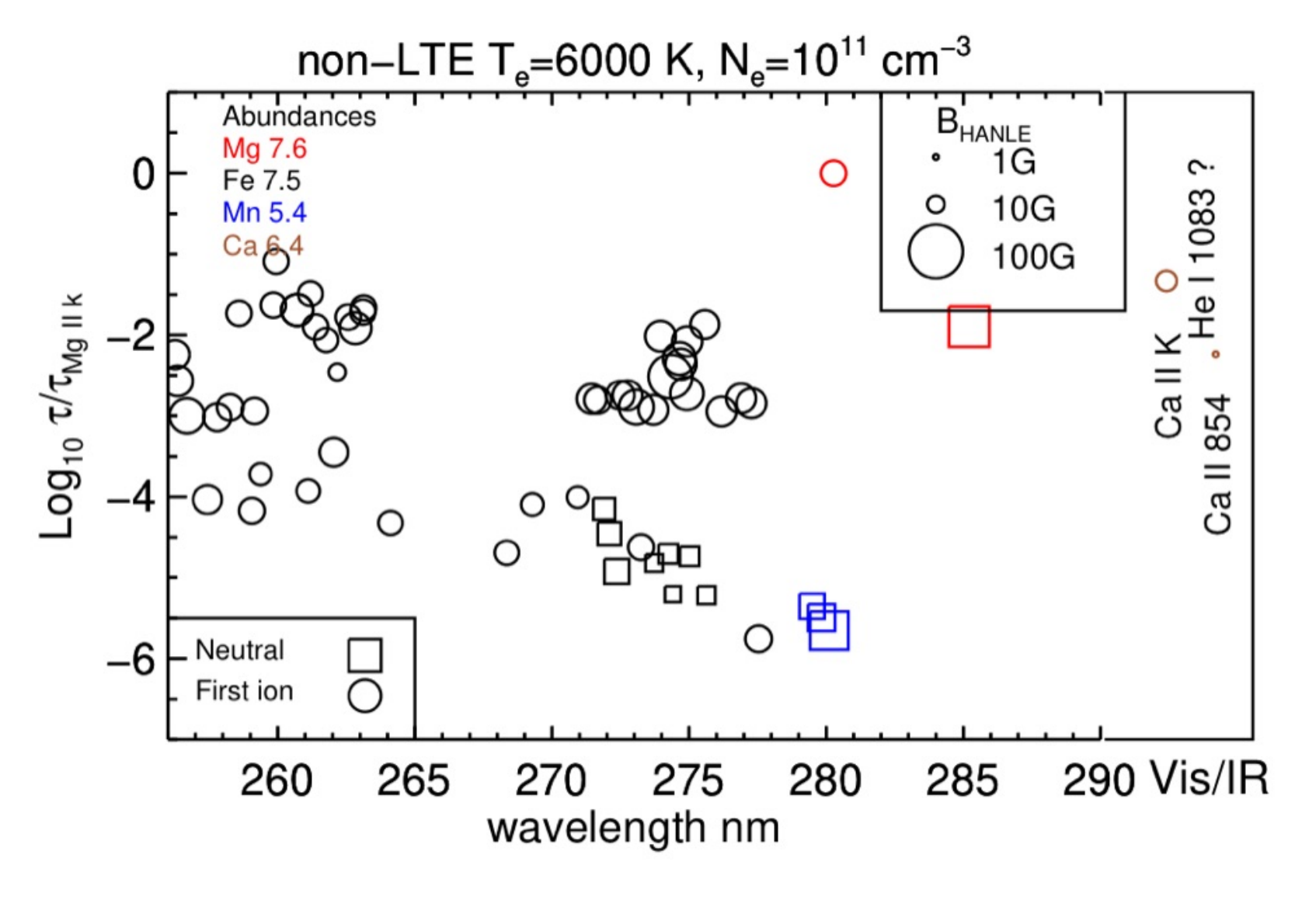}
\caption{Properties of 
spectral lines calculated using line center opacities as a function of wavelength are shown. 
Circles show lines of singly-charged ions, squares show lines of neutral species. Symbol sizes reflect critical Hanle field strengths. Relative optical depths of the lines are shown on the ordinate.  The \ion{Mn}{1} lines 
should include wing opacity
from the \ion{Mg}{2} line, but this was omitted here.  The \ion{Mn}{1} 
optical depths are thus a lower limit. 
The ``?''
for the He I 1083 nm line refers to the unknown relative optical
depth of this line because of
peculiarities in the formation of
this multiplet in the Sun \citep[e.g.][who discuss this multiplet in the context of a well-observed flare]{2015ApJ...814..100J}.}
\label{fig:hanle}
\end{figure}

The strongest line 
in the chromospheric visible spectrum is
\ion{Ca}{2} $K$ has just $1/20$ the opacity of the \ion{Mg}{2} $k$ line (see Figures
\ref{fig:contrib} and 
\ref{fig:hanle}). Also shown in Figure~\ref{fig:contrib} is the well-known  
line of \ion{He}{1} at 1083 nm which is present in active region plasma. However,  the line,
belonging to the triplet system, populated 
largely by recombination following EUV 
photoionization, 
tends to
form where coronal ionizing photons (and perhaps even hot particles) cannot penetrate the chromosphere.  Our 
non-LTE calculations (Figure~\ref{fig:contrib}) reveal that significant  opacity in the line is narrowly confined between surfaces of column mass 
well below  the height of
the NUV \ion{Mg}{2} lines, because of the depth of penetration of vacuum UV  radiation from the overlying corona in the 
continua of H and He.
When included along with the 
\ion{Ca}{2} resonance and 
infrared triplet lines, 
the optical- near UV regions accessible from the ground 
extend neither high enough, nor do they sample opacity with the fineness of UV lines (Figures~
\ref{fig:contrib}, 
\ref{fig:corrugated} and \ref{fig:hanle}.  Lines of  \ion{Mn}{1} are sufficiently close to the strong \ion{Mg}{2} lines for the wings of the latter to contribute to opacity, but 
this was not accounted for in the plot. Therefore heights  of formation of \ion{Mn}{1} lines are  lower limits. ).

The solar irradiance spectrum at vacuum UV wavelengths (below 200 nm, and above 91 nm below which the Lyman continuum absorbs line radiation),
including H L$\alpha$ and many  
chromospheric and transition region lines,
is shown in Figure~\ref{fig:alluv}.  All are substantially weaker than the \ion{Mg}{2} lines. The polarimetric measurements necessary to infer magnetic fields 
require large photon fluxes.
The VUV region 
is a factor of at least 10-100 times dimmer than the NUV.  
Also, the Zeeman signatures vary in proportion 
to the 
wavelengths observed. Further, most of the transition region lines are optically thin, so that 
there is no single ``depth of formation" of a given line. 
The exceptions are the Lyman lines of hydrogen
and 58.4 nm resonance line of He~I.   But 
using only this region of the spectrum would also fail to sample 
surfaces of multiple optical depths, and thus fail to 
address the need to span 
the anticipated corrugated isothermal and isobaric surfaces 
(Figure~\ref{fig:corrugated}), particularly in active regions, the sources of most solar eruptions.

\begin{figure}[ht]
\includegraphics[width=\linewidth,trim=10 10 20 10,clip]{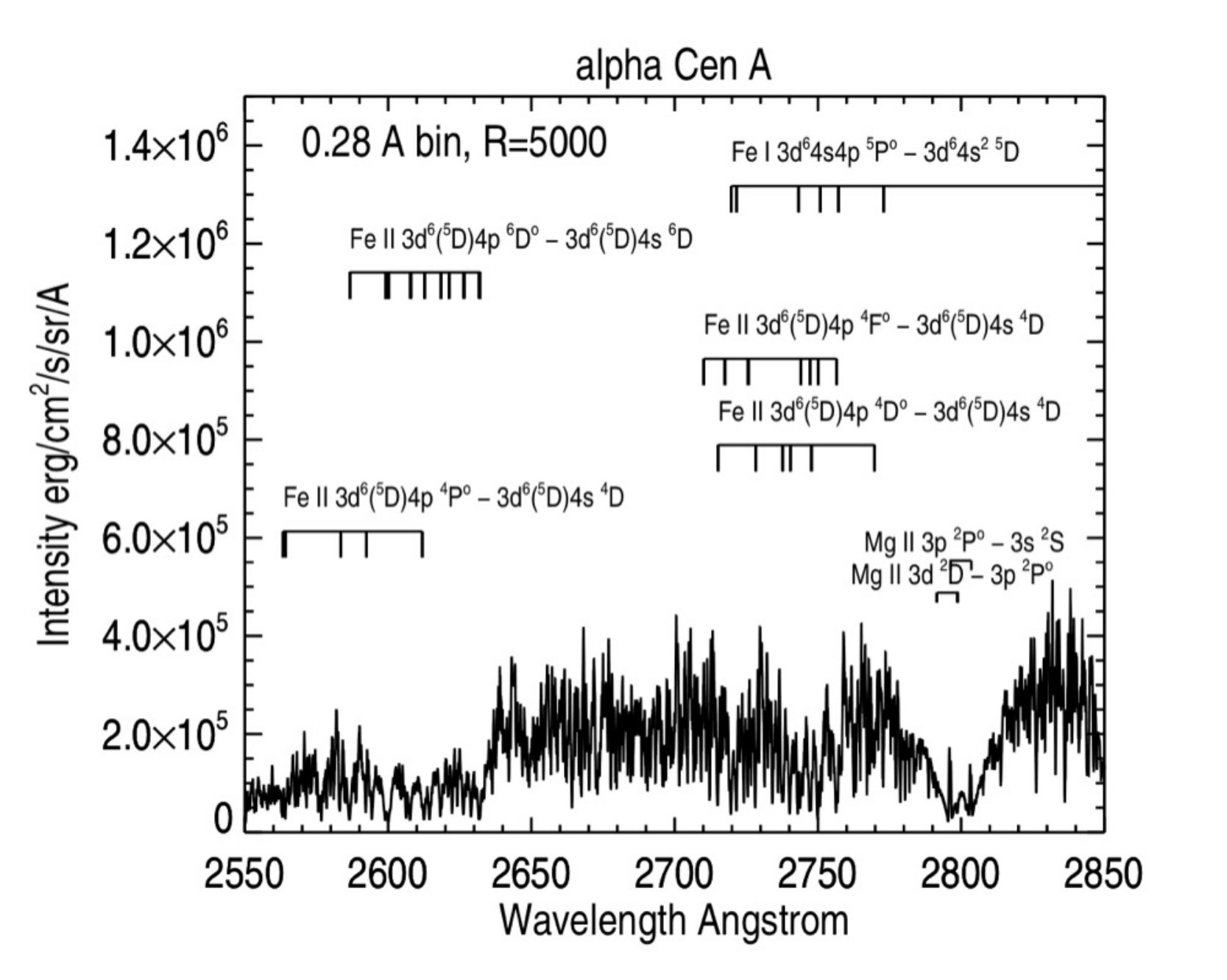}
\caption{A smoothed
UV spectrum 
of $\alpha$ Cen A, obtained by the Hubble Space Telescope, is displayed as a proxy for the 
mean solar intensity spectrum.  The large number of lines of \ion{Fe}{1} and \ion{Fe}{2} are indicated and annotated with their multiplet
terms.  The \ion{Mg}{2} spectrum is also 
annotated including the 
$3s-3p$ $h$ and $k$ lines, and the $3p-3d$ lines. Notice that the 
permitted \ion{Fe}{2} lines (shown with solid bars with downward ticks) consist of the sextet-D resonance lines $(4s-4p)$ near 2600 \AA, and quartet transitions at other wavelengths that occur between 
metastable (even parity) quartet lower levels and odd-parity quartets of the same transition array ($4s-4p)$. 
}
\label{fig:alphacen}
\end{figure}

In contrast, the NUV region (from say 200 to 310 nm, the atmospheric cutoff) is 
comparatively brighter,  containing 
lines (\ion{Mg}{2}) whose cores form within  lower transition region.  Compared 
with lines accessible from the ground, 
lines have smaller Zeeman signatures, but 
have increased
linear polarization generated by the higher  radiation anisotropies at UV wavelengths (much stronger limb-darkening). Thus, the 
Hanle signals will statistically be stronger than 
at visible wavelengths. Further, the region contains a 
plethora of lines of neutral and singly ionized iron
within the $4s-4p$  transition array, 
including  resonance lines near 260 nm (see Figure~\ref{fig:alphacen}).
Their complex spectra span and sample multiple depths across
the chromosphere, extending into  lower transition region plasma (see in particular the contours of unit optical depth surfaces for several lines in the lower right panel of Figure \ref{fig:corrugated}, and the plot of relative optical depths in Figure~\ref{fig:hanle}).

The richness of the \ion{Fe}{2} spectrum itself 
has been known for decades in astronomy  \citep[e.g.][]{Viotti+others1988}.  
Figure~\ref{fig:alphacen} identifies 
 lines and multiplets of {Fe} plotted above a smoothed Hubble Space Telescope spectrum of $\alpha$ Cen A, which is a good proxy for disk-averaged 
intensity spectrum of the Sun.  
(There is no disk-averaged solar spectrum of comparable quality). 
All of the strong transitions of iron lie 
near or below the \ion{Mg}{2} lines near 280 nm.  NUV spectra obtained 
at the solar limb
during the SKYLAB era revealed 
a spectacular array of \ion{Fe}{2} lines in emission against the darkness of space 
\citep{Doschek+others1977}.  Particularly strong in the limb spectra 
 (Figure~\ref{fig:skylab})
are multiplets in the sextet system near 259-263 nm, and quartet transitions in the 273 nm region. 
These \ion{Fe}{2} lines were readily detected  
up to 8" above the limb in the SKYLAB data, the limb itself is already
far higher than 
the photosphere.  Such emission is 
very likely associated with the kind of structures modeled in Figure~\ref{fig:corrugated}.   A downside of the richness of
the spectra of iron is the possibility of blended lines.  The important issues of blended spectra 
lines at NUV wavelengths and effects of missing opacity 
are addressed in the Appendix.  They are shown to be a minor issue, affecting only a few lines in the regions between 256 and 281 nm,
owing to the overwhelming strength of the \ion{Fe}{2} lines formed high enough in the 
chromosphere to aid in the diagnosis 
of magnetism there.   Between 281 and 310 
nm, there are just 10 ``wide" lines, i.e. strong enough to exhibit wing absorption profiles,  compared with 
38 listed by \cite{Moore+Tousey+Brown1982} 
between 256 and 281 nm. None of the lines above 281 nm form high enough in the Sun's atmosphere
to contribute significantly beyond  measurements
of the chromospheric magnetic field at shorter 
wavelengths.

Relative optical depths at the centers of the lines in this region, and ``Hanle field strengths'' 
are shown in 
Figure~\ref{fig:hanle}, including
lines of He, Mg, Ca, Mn and Fe. These Hanle field strengths (see equations \ref{eq:larmor} and 
\ref{eq:hanle} below)
are those near which the emergent  polarized spectra are strongly sensitive to 
magnetic field strength and direction, 
\citep[][see below]{Landi+Landolfi2004}.

\subsection{Spectral signatures of magnetic fields}

The Zeeman effect is well-known in laboratory and astrophysics as a characteristic splitting
in intensity ($I$) spectra. But in the Sun and stars, the Stokes parameters $I,Q,U$ and $V$ are used to measure magnetic fields in
commonly encountered situations
where the widths of the lines exceed the Zeeman splitting (this is assumed 
in the following discussion).  As noted above, Zeeman signals depend upon the quantity 
$\varepsilon = \omega_B/\Delta\omega_0$, where
$\omega_B$ is the Larmor frequency of the gyrating ion:
\begin{equation} \label{eq:larmor}
    \hbar \omega_B =  \mu_0 B.
\end{equation}
Here 
$\mu_0$ is the Bohr magneton,  and 
$B$ the magnetic field strength in G. 
$\Delta\omega_0$ is the 
line width (Doppler width for 
chromospheric and coronal lines). 
The amplitude of circular polarization 
measured by Stokes $V$ 
varies linearly with $\epsilon$, the linear polarization profiles $Q$ and $U$ have amplitudes varying as $\varepsilon^2$.  
While $B$ is measured in Gauss, it is important to remember that unless contributions from
non-magnetic regions to $I,Q,U$ and $V$ can be determined, the polarized components 
can be used to determine only the average flux density per unit area
in Mx~cm$^{-2}$, not field strength.

\tabblends

The Hanle effect has a quite different and complementary 
behavior independent of the Doppler width 
$\Delta\omega_0$.   It requires 
a pre-existing state of 
polarization induced 
by symmetry breaking effects such as anisotropic and/or polarized incident radiation.  When 
the radiation is anisotropic, such as radiation emerging into the 
upper chromosphere from the photosphere below, the atom 
develops atomic polarization 
(a particular imbalance of populations of magnetic substates called 
the atomic ``alignment'').   In the presence of a magnetic field,
these populations and associated 
atomic coherences (essentially 
representing the entanglement of neighboring atomic states) are altered by the magnetic field in the regime where
\begin{equation} \label{eq:hanle}
    g_{ji}\omega_B \mathcal{T}_j \sim 1,
\end{equation}
where $g_{ji}$ is the effective Land\'e $g-$factor for the transition from upper level $j$ to lower $i$, and $\mathcal{T}_j=
1 / \sum_iA_{ji} $ is the inverse lifetime of level $j$,
the sum taken over all lower levels $i$.  Figure~\ref{fig:hanle} shows spectral lines of Fe and Mg, their wavelengths, optical depths, with symbol sizes indicating the magnetic field for which equality holds  in  equation~(\ref{eq:hanle}), 
using data compiled in the NIST 
spectroscopy database. These Hanle field
strengths vary between roughly 
1 G and 70 G for the different lines, field strengths that are anticipated 
in quiet 
and active regions of the solar chromosphere. 
The data shown in Figure~\ref{fig:hanle} are listed in Table~\ref{tab:blends}, the final column of which which 
also addresses the potentially serious problem of spectral
line blends.   This problem is shown in the Appendix 
to be easily dealt with. 

In a classical picture, when $g_{ji}\omega_B \mathcal{T}_j \approx 1$, the 
ion on average gyrates through 1 radian 
before it emits photons sharing the common upper level. 
When $\omega_B \mathcal{T}_j \gg 1$, the ``strong-field limit'' of
the Hanle effect, the radiating 
ions exhibit many gyrations before
emitting a photon and all ``memory''
of the particular orientation of
the radiating ions is statistically averaged out.  This condition is the same as 
that found for coronal forbidden lines 
where $\mathcal{T}_j$ is typically between 1 and 0.01 seconds, not $10^{-8}$ seconds for 
the permitted lines studied here \citep{Casini+White+Judge2017}.  
The only signature of the magnetic field is in its direction projected on to the plane-of-the-sky, modulo 90$^o$.
If $\omega_B \mathcal{T}_i \ll 1$ 
(the weak field limit) the 
magnetic field enters as a small
perturbation in 
expressions for the emergent radiation and the magnetic effects cannot easily be observed.  But near the Hanle regime (equation~\ref{eq:hanle}) 
the magnetic field is encoded as
a rotation and de-polarization 
of the emergent linearly polarized radiation.  The symmetry broken by the mis-alignment between the directed radiation vector and 
magnetic field vector resolves
ambiguities associated with 
Zeeman measurements.   This effect may be useful for identifying tangential
discontinuities (changes of magnetic field direction not strength, across a flux surface).

In practice, 
both the Zeeman and Hanle effects can be used 
in a complementary
fashion to find 
signatures of magnetism in the Sun's atmosphere using such spectra
\citep{Trujillo2018,Ishikawa+others2021}.

Finally, we have investigated whether molecular transitions are significant in this spectral region, using the line list complied by R. Kurucz\footnote{ http://kurucz.harvard.edu/linelists.html} and the synthesis codes by \cite{Berdyugina+others2003}. In particular, we have found that OH transitions from the electronic A--X system are abundant in this spectral region, with stronger lines toward the red. They form below most chromospheric lines owing to chromospheric stratification and the rapid drop of molecular number density with gas pressure, but remarkably, can also form above deeper photospheric lines. A few moderately strong OH features may be visible at wavelengths between strong atomic lines in the quiet sun spectrum. They become especially important in models of sunspot umbrae. Numerous molecular CO lines in this spectral region are quite weak, but they are so many that the continuum level is effectively reduced by a few percent, with stronger lines toward the blue. If observed, both OH and CO lines are potentially interesting for sensing magnetic fields using both the Zeeman and Hanle effects 
\new{ 
\citep{2002A&A...385..701B,Berdyugina+others2003, Berdyugina+Fluri2004}, as well as temperature and pressure near the interface between the photosphere and chromosphere.
(The reader can refer to the paper by 
\citealp{2007A&A...461....1L} for
elementary theory of molecules in magnetic fields applied to solar 
conditions). }

We conclude that, \textit{based on known atomic and molecular data, there are many lines of \ion{Fe}{2} whose cores are not strongly blended}. The case of blended lines from 
sunspot umbrae requires 
further work outside of the scope of this paper.

\subsection{Diagnostic Sensitivity of Mg II \textit{k}}

\begin{figure}
    \centering
    \includegraphics[width=\linewidth]{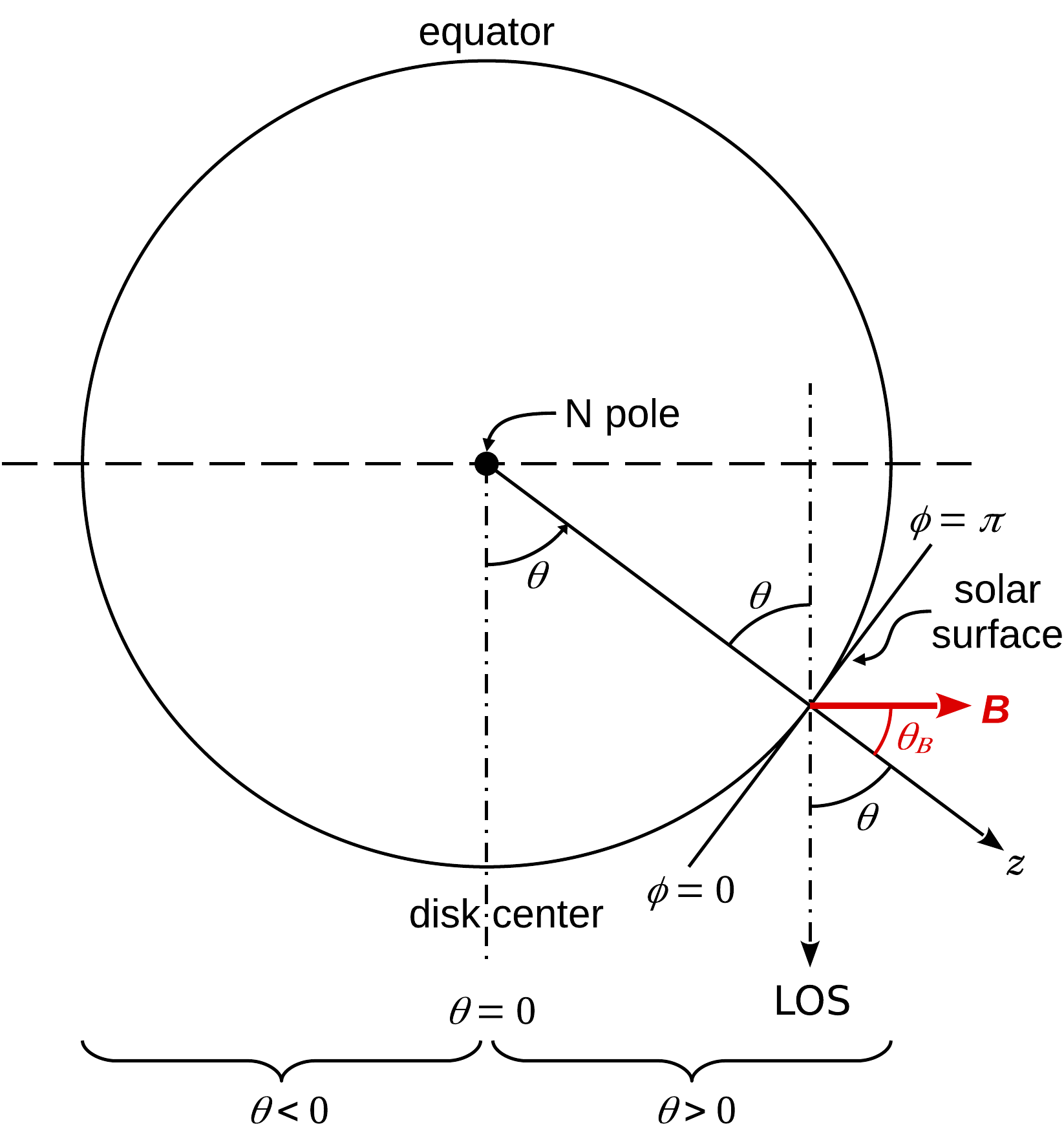}
    \caption{Observing geometry for uniform field computations with HanleRT used in the panels of Figure \ref{fig:heatmap}.  Note the convention for negative LOS $\theta$.}
    \label{fig:heatmap_geometry}
\end{figure}

\begin{figure*}
  \centering
  \includegraphics[width=0.9\textwidth]{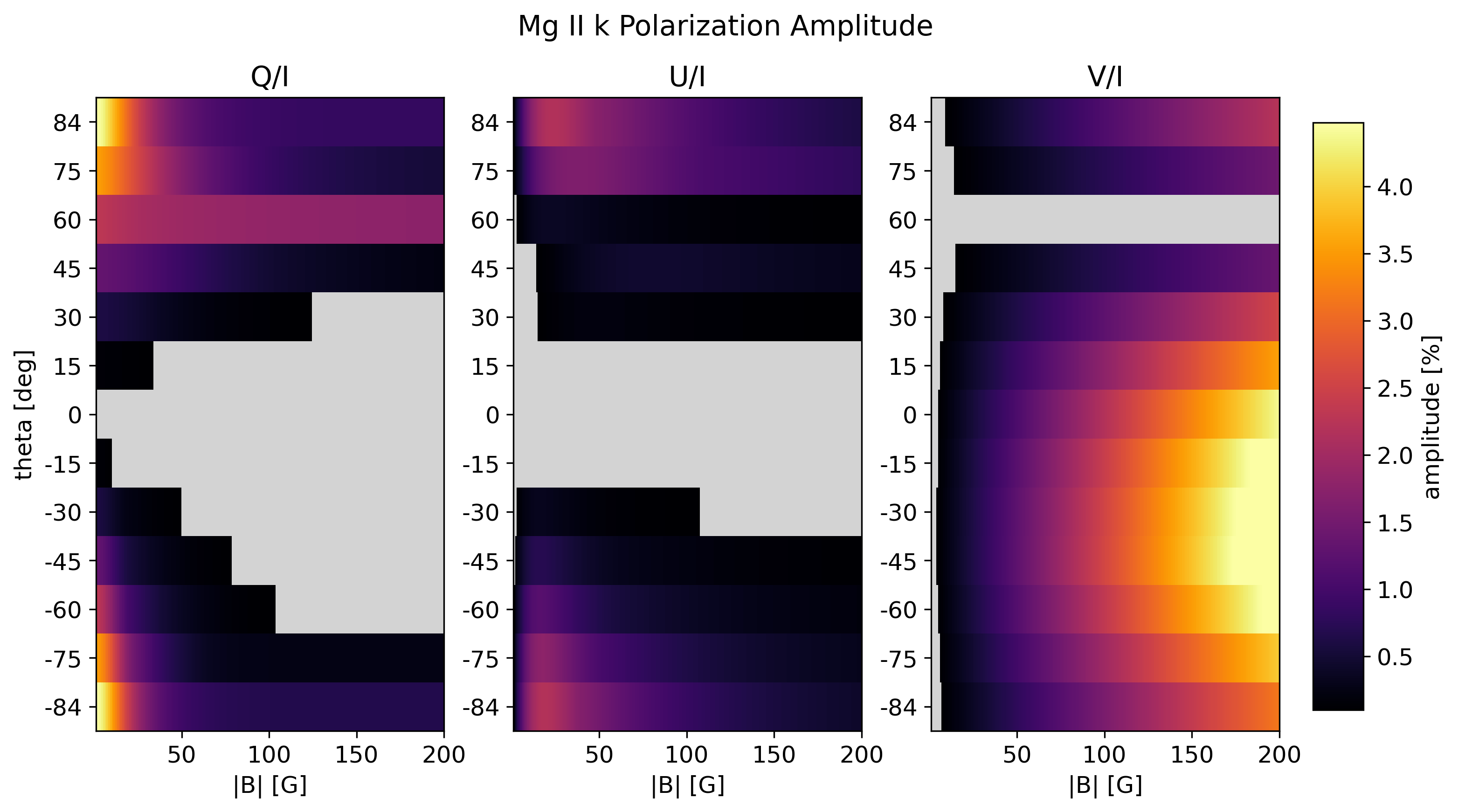}
  \includegraphics[width=0.9\textwidth]{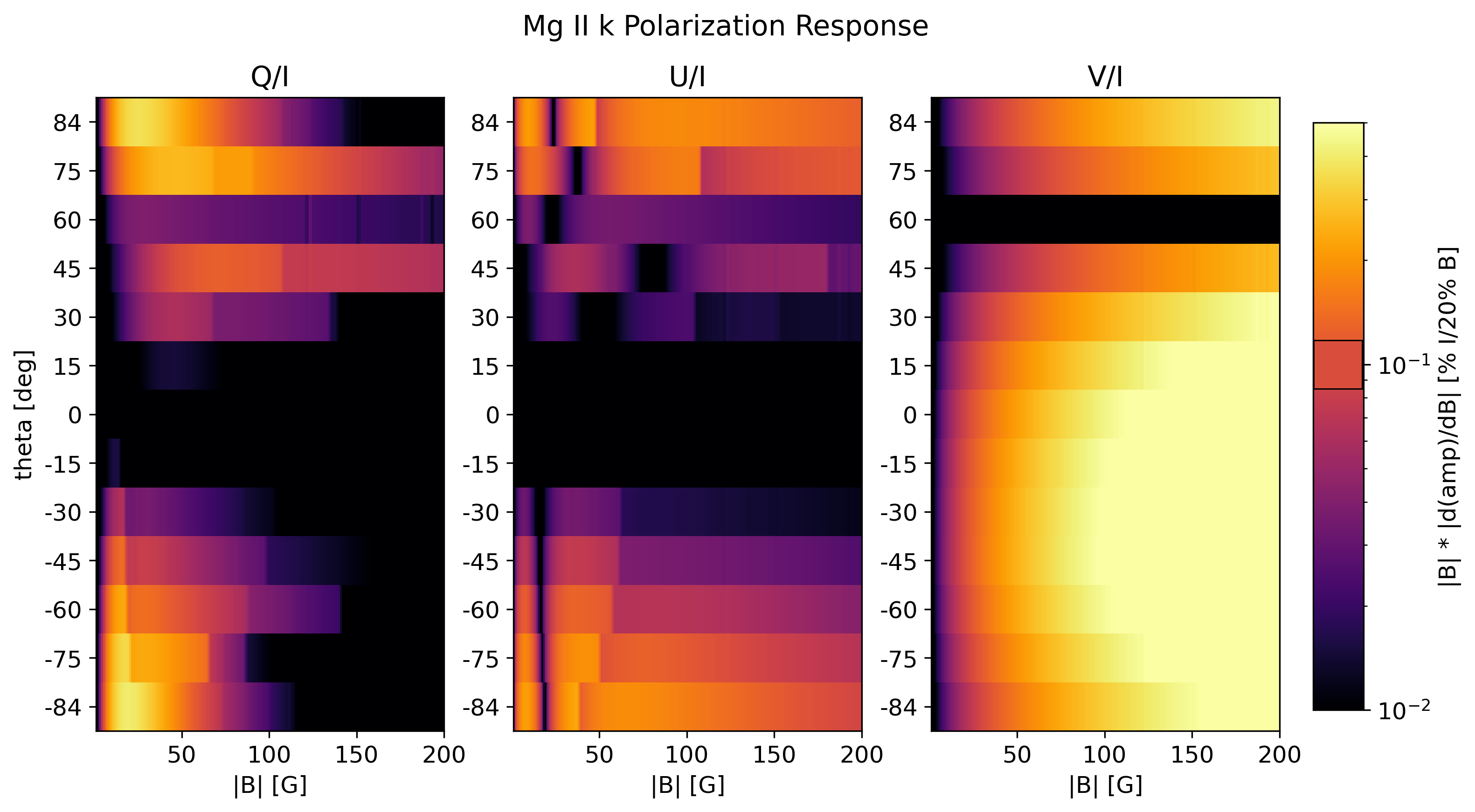}
  \caption{
  \emph{Top:} ``Heat Map'' of the \ion{Mg}{2} k core polarization amplitude for a uniform magnetic field inclined at $\theta_B = -30\degrees$ from vertical.  Gray regions are where the amplitude falls below 0.1\% of the intensity, here taken as a typical instrumental SNR. 
  \emph{Bottom:} Derivative of polarization amplitude data from the top panel with respect to magnetic field to produce a ``heat map'' of the response.  The color scale gives the relative change in amplitude
    of the spectral feature in response to a 20\% change in $|B|$.
    Regions of the map brighter than the 0.1\% value on the color
    scale (highlighted by black box) indicate an observing geometry that could infer $|B|$ to within 20\%
    assuming an instrumental polarization SNR of 0.1\% of the
    intensity.  See text for further details.}
    \label{fig:heatmap}
\end{figure*}

In Figures~\ref{fig:contrib} and
\ref{fig:corrugated} 
we saw
how the \ion{Mg}{2} $k$ line 
has contributions from plasma
that is highest in the Sun's
atmosphere.   The $h$ line is a factor of two less opaque, but it has a higher Land\'e g-factor than $k$ and so 
is more sensitive to the Zeeman effect
\citep{Ishikawa+others2021}.   However, its upper $J=1/2$ level
is unpolarizable and so it can have no
Hanle effect in the line core. 

Both the Zeeman and Hanle effects
can and should be used to probe
chromospheric magnetism.  The Zeeman effect produces the largest
polarimetric signals when fields are relatively strong (average flux densities of a few hundred Mx~cm$^{-2}$  were measured using the weak field limit applied to the Mg lines by \citealp{Ishikawa+others2021}). However,
for weaker fields, the Hanle effect becomes the dominant source of polarized light variations in response to the solar magnetic fields.  Following early theoretical work 
by \cite{2012ApJ...750L..11B},
the scattering polarization, Hanle and Zeeman effects across the $h$ and $k$ lines was investigated  in semi-empirical 1D
models \citep{Alsina+others2016, delPinoAleman:2016,delPinoAleman:2020}. The Hanle field of the $k$ line is 23 G (see Figure~\ref{fig:hanle}).  These authors demonstrated that, for
field strengths of the order of 10 G, measurable variations of the
order of $\sim$1\% in the fractional linear polarization of \ion{Mg}{2}
$k$ are produced by the Hanle effect. 
Here, we extend this work with a larger
grid of LOS angles and magnetic field strengths (see Figure \ref{fig:heatmap_geometry} for the geometry) and show polarization amplitude and magnetic
response ``heat maps'' in Figure \ref{fig:heatmap}.  As in
\citet{delPinoAleman:2020} we used HanleRT to compute the intensity and polarization from a \ion{Mg}{2}
3-term atom including PRD effects in atmosphere C of \citet{Fontenla+Avrett+Loeser1993}.  We
computed profiles for a grid
of uniform magnetic field vectors inclined at $\theta_B = -30\degrees$ from
vertical, with fixed azimuth,  and with $|B|$ ranging from 1 to 200 G at 1 G intervals.  These values of $|B|$
are about 20 and 10 times smaller and larger, respectively, than the Hanle 
field.
We computed the emergent Stokes profiles for 13 LOS viewing angles $\theta$ ranging
from $+75\degrees{}$ to $-75\degrees{}$ at 15\degrees{} intervals, as well as
$\mu =\cos(\theta) = 0.1$, which corresponds to $\theta \sim \pm84.5$\degrees{} or a limb distance of about 5$''$.
We examine the relative variation in the amplitude of the fractional
polarization (hereafter simply ``response``), $|B| \cdot dA_i/dB$, where $i$ indicates the 
polarization $Q$, $U$, or $V$, and $A_i$ is the amplitude,
$A_i = ({\rm max}(i) - {\rm min}(i)) / I \cdot 100\%$, within the \ion{Mg}{2} k core region.  
The top panel of \ref{fig:heatmap} shows the amplitude $A_i$, 
while the bottom panel shows the response with respect to magnetic field over the full range of $|B|$ and LOS $\theta$.  
The units of the response are $\% I / \% B$, and we have scaled the data such that the denominator represents a 20\%
change in $|B|$.  Under this normalization, a response of 1 corresponds to a 1\% amplitude change for a 20\% change
in $|B|$. 

In order to assess the practical
application of 
these calculations, we consider % note: noise was not added to the calculation
noise at the level of 
$\sim 0.1 \%$ of intensity ($10^{-3} I$). Those regions in the response heat map brighter than
that value on the color scale are where the instrument would be able
to infer components of the magnetic field to within 20\%.
We conclude that we would be able to
reasonably estimate this vector orientation for magnetic field
strengths of $\sim$5 to 50 G at observation angles from about
$\theta = \pm 45\degrees$ to the limb due to the Hanle effect in $Q$ and/or $U$,
and stronger line-of-sight fields at nearly any observation angle from the Zeeman
effect in Stokes $V$.  The largest sensitivity in $Q$ and $U$ occurs for
observations near the limb and field strengths below the critical
Hanle field.  (Dark vertical strips in the response of Stokes $U$ are due
to local maxima in the amplitude curve near the critical Hanle field
for \ion{Mg}{2} $k$; compare to the top panel of Figure \ref{fig:heatmap}).  The $V$ amplitude is constant and zero at $\theta =
60\degrees$ as this is the angle where the magnetic field vector is orthogonal
to the line-of-sight.  Measurable linear polarization is produced at that geometry, 
however its variation with magnetic field is too low to be used as a diagnostic.  
``Heat map'' diagrams for other field inclination angles
are qualitatively similar, with the regions of maximum response 
due to the Hanle and Zeeman effects shifting in relation to the field \& LOS geometry.  

We conclude that diagnosing magnetic field from the Hanle
effect in \ion{Mg}{2} $k$ is feasible for a large range of geometries
and with achievable signal-to-noise ratios.  The range of field
strengths to which the Hanle effect is
sensitive complement those higher 
field strengths ($|B| \ge 50$G)
to which the Zeeman effect is sensitive, for
the case of the \ion{Mg}{2} $k$ line.

Additional work on the Hanle effect in
the lines of \ion{Fe}{2} is underway
and will be reported elsewhere. 
Importantly, there is at least one line 
(at 2744.01 \AA) with a Hanle field 
above 60 G, and several where the
critical field exceeds that of the \ion{Mg}{2} $k$ line.

\subsection{Addressing the ``corrugation problem''}

Figure~\ref{fig:corrugated} shows 
how state-of-the-art numerical models
lead to contours in optical depth 
which can be far from horizontal.
The free energy in these MHD simulations generated by convection and emerging flux leads to highly dynamic 
evolution in the lower density regions of most interest to us at the top of the chromosphere.   The geometric distortions of the $\tau=1$ surfaces in these 
tenuous regions naturally reflect 
the low inertia of the plasma, subjected to  
Lorentz, pressure gradient and gravitational forces, evolving 
in response to the forcing from below.
But, as noted earlier, the extreme corrugations of order 10 Mm are not negligible compared with the scale of the emerging flux system itself. The question then arises,  how can we infer 
magnetic field at a constant geometric height, if we are to
apply mathematical techniques 
to estimate the free energy and topology within the corona?  Progress can be anticipated without such 
techniques, by direct comparison of numerical 
prediction and measurements.  But this question
must be addressed if we are to measure the development of
free energy in the overlying corona in time,
in general.

Let us assume that the calculations 
are a reasonable, ``first order'' 
approximation to the kind of geometric distortions of the $\tau=1$ surfaces of sets of lines.   The advantage of observing the 259-281 nm region is that we know the order of
the opacities of each line in a sequence: the \ion{Mg}{2} lines 
are the most opaque, next come the 
sextet system transitions of \ion{Fe}{2} and then the quartet
systems, each with its spread of
oscillator strengths.  Thus, given
a calculation, we can (as in Figure~\ref{fig:corrugated}) find the 
$\tau=1$ surfaces.  We could 
compute Stokes profiles and associate a magnetic field from such 
synthetic data, knowing the heights
of each line's formation. 

But this is not the situation we
face with the remotely-sensed 
Stokes data, we have no knowledge
other than the Stokes profiles of many lines, on a two dimensional 
geometric image of the Sun at each
wavelength.  Fortunately, this problem fits well into the picture
of machine learning algorithms
\citep[e.g.][]{Bobra+Mason}.  Synthetic data can be used to train 
an algorithm to find level geometric surfaces \citep{2019A&A...626A.102A} of direct use for mathematics to derive quantities of
interest.  The large coverage 
of optical depths of the 256-281 nm
region and the variety of Hanle field
strengths strongly suggest that, if
the models are sufficiently complete
to represent the atmospheric dynamics
and temperature structure, then 
such an algorithm should have 
a non-zero probability of yielding 
the desired results.  The combined 
use of Hanle and Zeeman effects enhances this probability, at least in principle.  We are exploring 
ways in which this program of
research might be implemented.

%%%%%%%%%%%%%%%%%%%%%%%%%%%%%%%%%%%%%

\section{Discussion}

In this paper we have examined the general problem of identifying the sources of magnetic energy within coronal plasma, which leads to flaring, plasmoid and CME ejections.   By a process of
elimination, we arrived at
a wavelength range between 
2560 and 2810 \AA{},  which appears to be the most promising range in which to
measure conditions at the coronal base.   Our conclusions build upon 
the important study of \citet{Trujillo+others2017}, not only
by seeking answers to questions concerning the magnetic free energy and topology, aimed at 
understanding the origins of 
phenomena responsible for Space Weather. 
Our conclusions also differ 
by proposing 
a potentially ideal,
relatively narrow spectral region 
between 256 and 281 nm, \new{expanding 
significantly the 279.2-280.7 
region} selected by
the CLASP2 mission, 
to serve 
as a basis for future, more targeted research.

To achieve measurements in this region clearly will require a space-based platform for spectropolarimetry.   These conclusions are based upon 
several criteria:  The spectrum must be bright enough; spectral features must extend as high as possible; the spectral features must contain information on magnetic fields through both the Zeeman and Hanle effects; the spectra must be demonstrably unblended;
lines with multiple 
opacities are needed to
span across the 
corrugated 
surfaces that are present 
at the coronal base. 

Thus we suggest that 
a novel spectropolarimeter,
building upon a heritage 
from SKYLAB, SMM (UVSP), SoHO (SUMER), IRIS, CLASP and CLASP2, 
be considered for flight to
address the problems of interest to science and 
society.   Such a mission 
would complement
the unique capabilties of
the Daniel K. Inouye 
Solar Telescope which can observe
important chromospheric 
lines (include all those shown in the top row of Figure~\ref{fig:contrib}) from the ground, at a sub-arcsecond spatial
resolution and/or high cadence.   

Finally, it is important to note that 
the CLASP2 measurements of 
\cite{Ishikawa+others2021} achieved signal-to-noise ratios of over 1000 in the \ion{Mg}{2} line cores.   This demonstrates that a modest, D=25 cm telescope 
operating at NUV wavelengths 
is sufficient for delivering 
the weak signals required for diagnosis of magnetic fields, over active regions at least, where the magnetic flux density  exceeds 100 Mx~cm$^{-2}$. 

\acknowledgments
Thanks to Roberto Casini and Alfred de Wijn for the many fruitful discussions regarding the Hanle effect and requirements to achieve new results using diverse spectral lines.
NCAR is funded by the National Science Foundation. 
PGJ is grateful to the NSF 
for funding his research 
at a national center without 
any specific goal.  The authors thank an anonymous referee for their helpful comments.

\appendix

\section*{Line blends and missing opacity}
\label{sec:spectral_syn}

The NUV region is  well-known 
as a crowded region of the solar spectrum, and also 
a region where some opacity
appears to be ``missing''.
The effects of this missing opacity 
are examined below using 
comparisons of detailed 
simulations and observations. It can be significant at certain wavelengths, but does it not
affect the cores of strong  chromospheric lines.  
(See also discussions by \citealp{Fontenla+others2011,Peterson+others2017}).
Line blending is however a  concern.  

We have addressed blending using data from the Hubble Space Telescope, from the SO82B spectrograph on
SKYLAB, and using sophisticated radiative transfer models.  
The models are based on Local Thermodynamic Equilibrium (LTE) radiative transfer models computed using TURBOSPECTRUM \citep{plez2012}, and adopting a custom calculated Solar 1D, spherical MARCS model atmosphere \citep{gustafsson2008}. A complete linelist in the corresponding spectral region was adopted in the computations from the VALD database\footnote{http://vald.astro.uu.se/} \citep{Ryabchikova+others2011}. The linelist includes up-to-date atomic data, including line-broadening parameters from collisions with neutral hydrogen atoms, calculated  \textit{ab-initio} following \citet{ABO} for all neutral atoms including dominant \ion{Mg}{2} lines and iron group elements.

\begin{figure*}[ht]
\centering
\includegraphics[width=.7\linewidth]{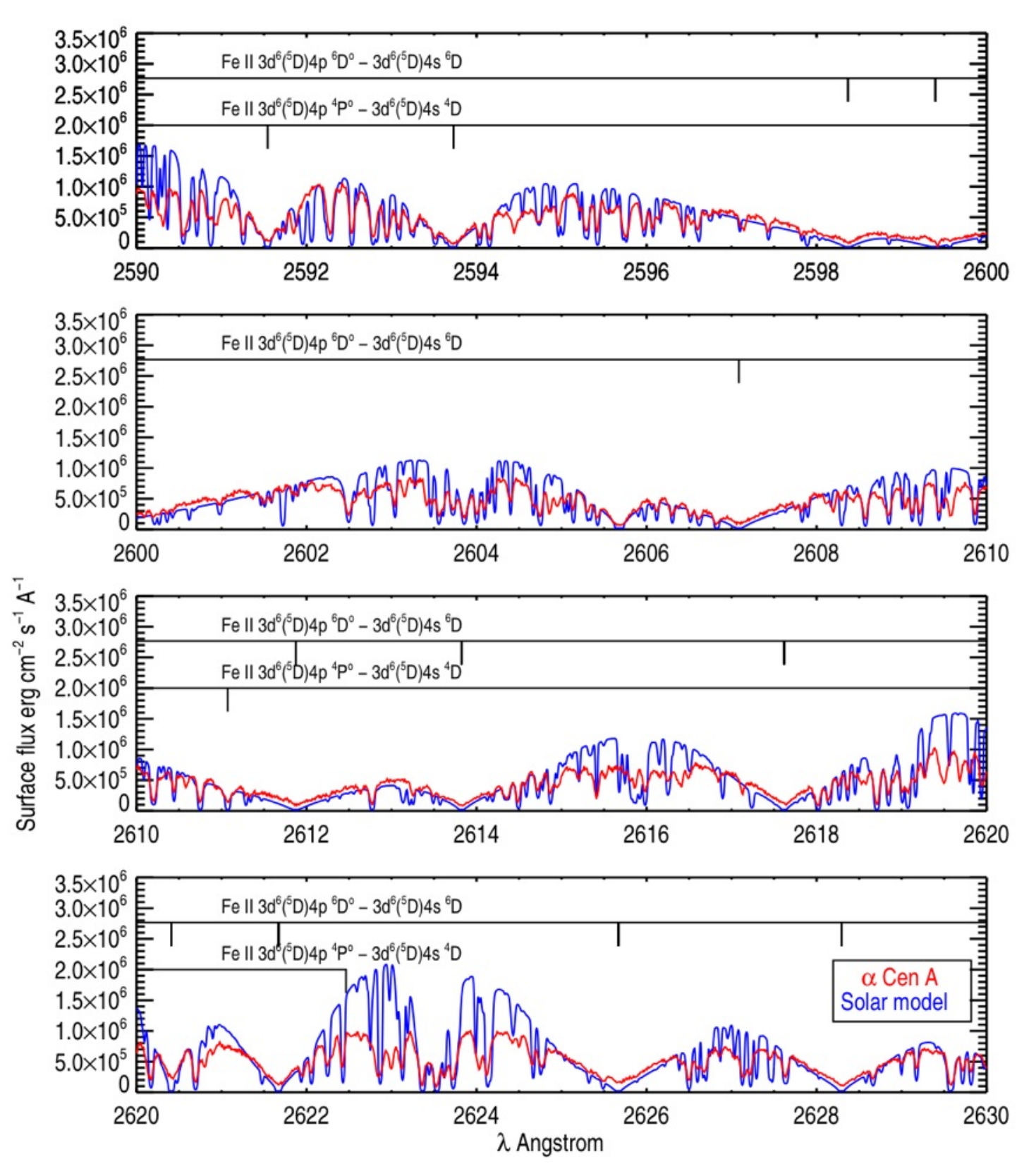}
\caption{Flux spectra computed for a Solar atmospheric model {(blue)} using 1D, LTE radiative transfer calculations (see Section\,\ref{sec:spectral_syn}) are shown along with spectra of the Sun-like star
$\alpha$ Cen A {(red)} obtained with the STIS instrument on the HST.   The data are a composite of several exposures from the E230H grating, assembled and processed by \cite{Ayres2014}.  The STIS data were shifted according to the redshift of 22.4 km~s$^{-1}$, and converted to flux at the stellar surface using the angular diameter of 9.09 milli-arcsec. The spectra are compared with no adjustment
of flux or wavelength scale.  The qualitative agreement is remarkable.
The units are \AA{} (wavelength) and 
erg~cm$^{-2}$~s$^{-1}$~\AA$^{-1}$ (flux, $F=\pi \overline I$,
where $\overline I$ is the mean disk
intensity). }
\label{fig:acena}
\end{figure*}

A comparison of high resolution Hubble
spectra of $\alpha$ Cen A with the models 
is informative.   A typical example in
an interesting region of the NUV spectrum is
shown in Figure~\ref{fig:acena}.  The blue line represents the flux spectrum 
computed for the Sun, the red line 
the spectrum of $\alpha$ Cen A.
The flux densities shown 
are computed at the respective stellar surfaces, with no ad-hoc adjustments 
of either scale.

The comparison is remarkable, particularly in the cores of some of the strong lines, most of which belong to \ion{Fe}{2}.  Wavelengths are evident (near 2623 \AA{} for example)
where opacity appears to be missing
(blue line above red line), but in the \ion{Fe}{2} lines identified in
the figure there is no evidence for
missing opacity, if anything, the observed line cores in red lie above the blue line, indicating emission
from the star's chromosphere. 

\begin{figure*}[ht]
\centering
\includegraphics[width=.7\linewidth]{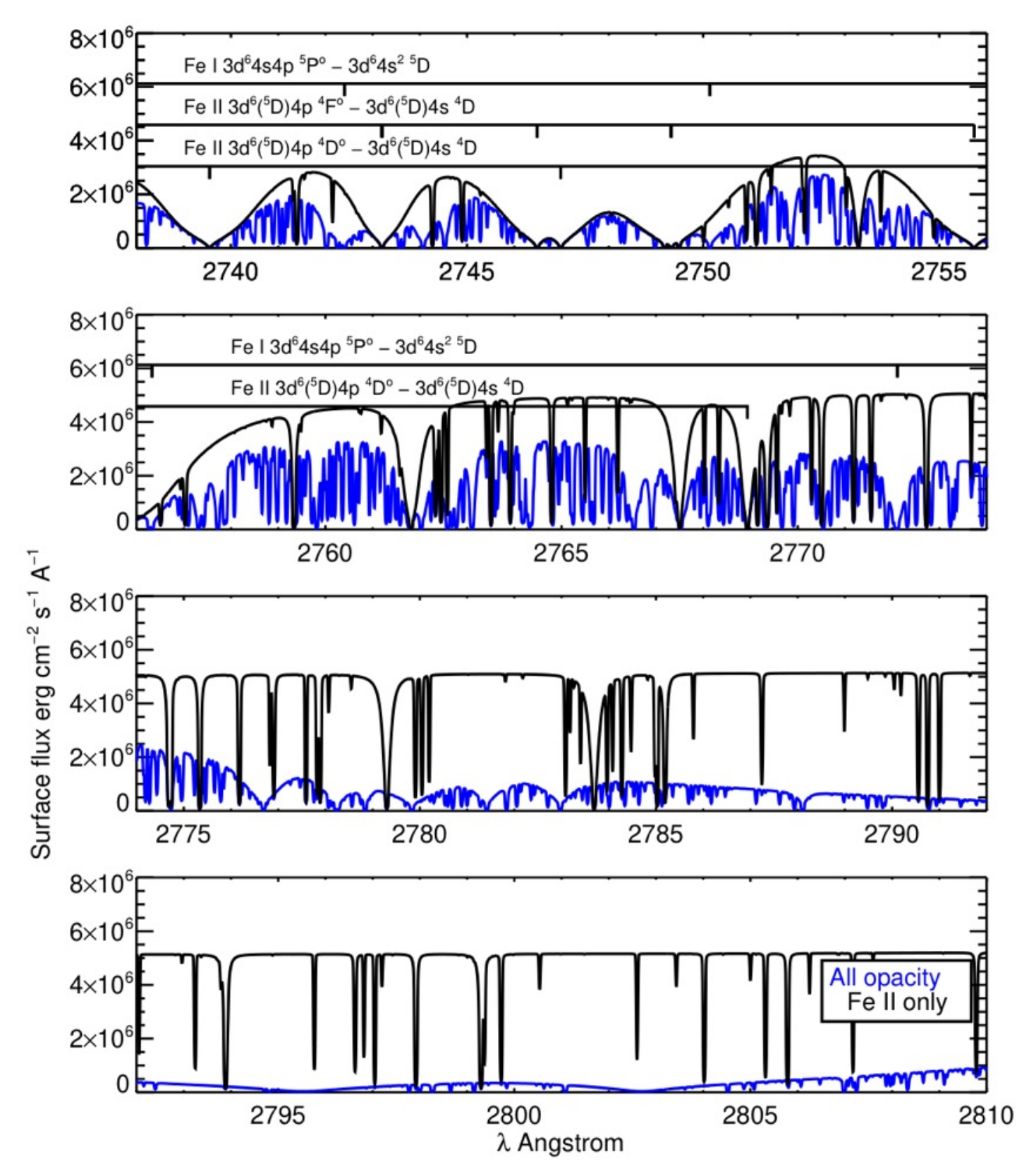}
\caption{ Two calculations 
are shown for the solar NUV spectrum: one (black) is made with the full set of lines, and 
the other (red) is made 
where lines of \ion{Fe}{2}
are the only sources of 
opacity.  Strong lines from
the quartet system of \ion{Fe}{2} are annotated, along with  lines of \ion{Fe}{1} whose opacity is much smaller. The cores of the strong lines of \ion{Fe}{2} are dominated by \ion{Fe}{2} itself, the region above $\approx$ 265 nm is, aside from narrow lines, dominated by wings of the \ion{Mg}{2} $h$ and $k$ lines.}
\label{fig:blends}
\end{figure*}

The  
model calculations also permit us to identify 
possible blends (Figure~\ref{fig:blends}). 
This figure shows 
two computations, black is
the full model calculation 
(also shown in blue in Figure~\ref{fig:acena}), and in this figure blue is a calculation using opacity only from the \ion{Fe}{2} lines.  When 
significant differences between black and blue lines exist, these are wavelengths where opacities are present in lines other than \ion{Fe}{2}.  Conversely,
when they agree, the \ion{Fe}{2} spectrum dominates and these are essentially \textit{unblended} wavelengths.  
As noted in the main text, the wavelength region above 281 and below 310 nm does not
contain strong lines formed in the middle-high
chromosphere, therefore these are not shown here.  Above 281 nm the lack of such lines
makes this a less desirable region because
blending becomes a bigger problem.

\begin{figure*}[ht]
\centering
\includegraphics[width=.45\linewidth]{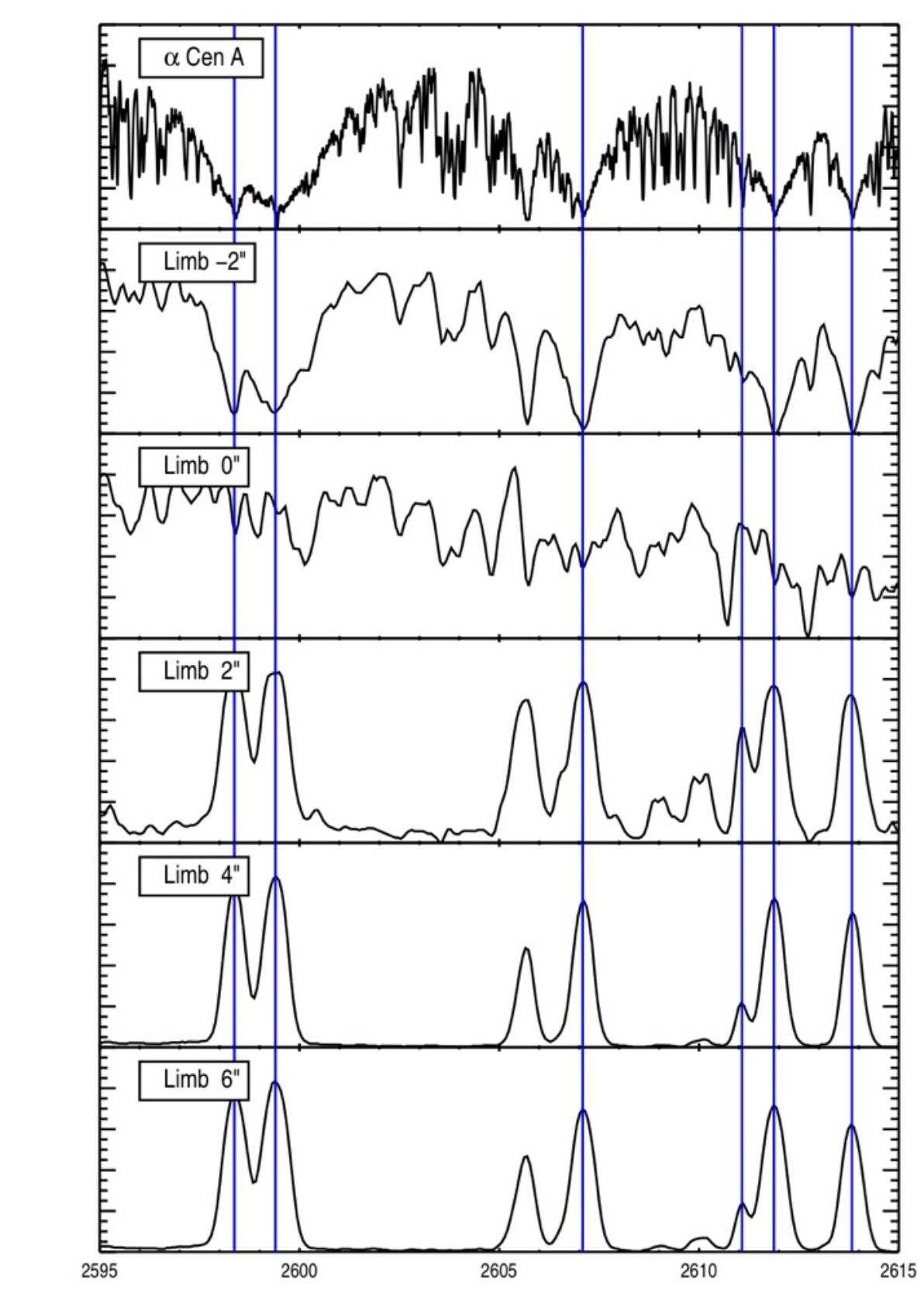}
\includegraphics[width=.45\linewidth]{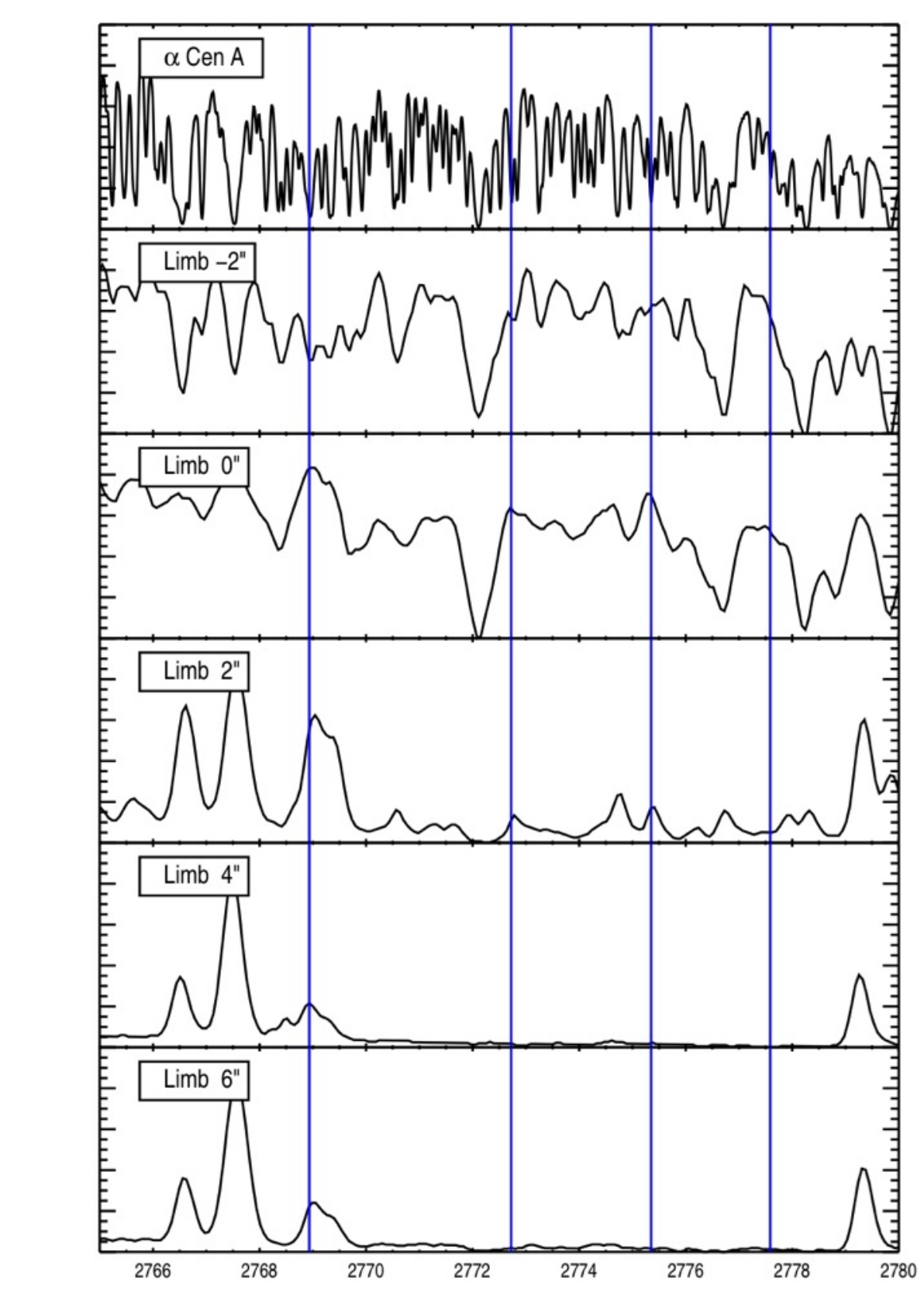}
\caption{The left panel shows the spectral region
close to the sextet resonance lines of \ion{Fe}{2} is shown 
from $\alpha$ Cen A (top),
and 5 spectra representative of the spectra as a function of position relative to the solar limb.  The right panel shows the 2765-2780 \AA{} region.  Blue lines mark wavelengths of \ion{Fe}{2} transitions.  The three other lines seen in emission in the right panel are from \ion{Cr}{2} at 
2766.52 \AA{}, and questionable identifications of 
\ion{Fe}{2} 2767.50, a transition 
between high lying states, and 
of \ion{Mg}{1} at 2779.8 \AA{}  \citep{Doschek+others1977}.
}
\label{fig:skylab}
\end{figure*}

Lastly we can, to some extent, check for blends independently of 
the model computations. \cite{Stencel+Rutten1980} were able to find weak lines in the wings of the \ion{Ca}{2} $H$ and $K$ lines looking to data obtained near the solar limb. 
In Figure~\ref{fig:skylab} 
we show (with an arbitrary flux calibration), spectra of $\alpha$ Cen A with spectra from the 
SO82B spectrograph on SKYLAB.  The latter are 
taken from the scans 
across a quiet region of solar limb obtained on August 27 1973, the slit being tangential to the limb and stepped by 2" between 
pointings.  
These are the photographic data analyzed by \cite{Doschek+others1977}.
The ``intensities''
plotted here are  photographic densities.  But these densities suffice for us
to find significant 
blends, which will potentially be revealed by changes in the spectra as the spectrograph slit was repointed from 2" inside the limb to 6" outside, in 2" steps. 
Different behaviors are seen depending on the line 
of interest.   In the wavelength
range from 2595 to 2615 \AA{} (left panel, showing resonance lines of the sextet system) there is
no evidence of blended features.  But the right panel of Figure~\ref{fig:skylab}
shows a spectral region containing lines of the quartet system of \ion{Fe}{2}.   In contrast to the resonance sextet transitions, here we find that essentially no \ion{Fe}{2} lines are strong enough to be unblended. However, a plot of similar transitions (not shown) between 2750 and 2765 \AA{} has stronger lines at 2756 and 2762 \AA{} which are not strongly blended.
These conclusions have been verified 
using NRL report 8653 \citep{Moore+Tousey+Brown1982}, from which the last column of Table~\ref{tab:blends} has been 
constructed.  There are only four 
lines which are classified as
severely blended of all the quartets, sextets of \ion{Fe}{2} 
and quintets of \ion{Fe}{1}.

\bibliography{biblio}
\bibliographystyle{aasjournal}

\end{document}